\documentclass[english]{article}
\pdfoutput=1
\usepackage{amsmath}
\usepackage{amssymb}
\usepackage{epsfig,cite}
\def\FIGURE#1{\begin{figure}[ht]#1\end{figure}}
\makeatletter

\usepackage{babel}

\def\be#1{\begin{equation}\label{#1}}
           
          \def\ee{\end{equation}}
\def\equ#1{~(\ref{#1})}

\begin{document}
\begin{titlepage}
\thispagestyle{empty}

\bigskip

\begin{center}
\noindent{\Large \textbf
{A Transfer Matrix Method for Resonances in
Randall-Sundrum Models III: An analytical comparison.}}\\

\vspace{0,5cm}

\noindent{G. Alencar ${}^{a}$\footnote{e-mail: geovamaciel@gmail.com }, R. R. Landim ${}^{b}$, M. O. Tahim ${}^{a}$ and R.N. Costa Filho ${}^{b}$}

\vspace{0,5cm}
 
{\it ${}^a$Universidade Estadual do Cear\'a, Faculdade de Educa\c c\~ao, Ci\^encias e Letras do Sert\~ao Central- 
R. Epitcio Pessoa, 2554, 63.900-000  Quixad\'{a}, Cear\'{a},  Brazil.
 
\vspace{0.2cm}
 }
 {\it ${}^b$Departamento de F\'{\i}sica, Universidade Federal do Cear\'{a}-
Caixa Postal 6030, Campus do Pici, 60455-760, Fortaleza, Cear\'{a}, Brazil. 
 }

\end{center}

\vspace{0.3cm}

\begin{abstract}

The transfer matrix method is used to analyze resonances in Randall-Sundrum models.
Although it has successfully been used previously by us we provide here a comparison between the numerical 
and analytical models. To reach this we first find new exact solution for the scalar, gauge, Kalb-Ramond and $q-$form fields. Them we calculate numerically the 
resonances by the transfer matrix method and compare with the analytical result. For completeness, this is done for models with and without the dilaton coupling. The results 
show a perfect agreement between the analytical and numerical methods.

\end{abstract}
\end{titlepage}

\section{Introduction}

 One of the main models treating aspects of physics of extra dimensions is the Randall-Sundrum model \cite{Randall:1999vf, Randall:1999ee}. This model gives a possible solution to the hierarchy problem and tell us how gravity is trapped in our four dimensional world. However, the lack of a more physical application have posed some challenges to this model through the past years. One of its problems is the appearance of spacetime singularities due to the presence of infinitely thin membranes. In such case, more realistic models based on smooth solutions representing the  four-dimensional membrane embedded in a higher dimensional spacetime is needed. In special cases, where the membranes are generated by scalar fields, its is very simple to obtain solutions through the superpotential method. In particular, several membrane types were considered: those generated by models with more than one scalar fields,  deformed membranes and so on\cite{Bazeia:2008zx, Afonso:2007gc, Bazeia:2005hu, Bazeia:2004dh, Fonseca:
2011ep, 
Fonseca:2010va, Yang:2012hu }. Beyond that, the analysis of how  the localization is made for various fields (tensors and spinors) have been understood \cite{Almeida:2009jc, Sousa:2012nx, Liu:2011zy, Christiansen:2010aj, Castro:2010au, Zhao:2010mk, Liu:2010pj, Chumbes:2011zt}. Other aspects include discussions about the tensions of membranes \cite{HoffdaSilva:2011bd,Abdalla:2010sz }. 

Another interesting problem is related to understand details of the interactions between the membrane and several particles that are not the usual zero modes. In other words, the approach to this problem considers the computational aspects of resonances for various models. In all of the models the massive spectrum is determined by a Schr\"odinger like equation with a potential that falls  to zero at infinity. The spectrum is not discrete and we have a ill defined effective action. Despite of this, just like in the case of semiconductors heterostructures, there is the possibility of appearance of resonances. This analysis have been done extensively in the literature \cite{Bazeia:2005hu,Liu:2009ve,Zhao:2009ja,Liang:2009zzf,Zhao:2010mk,Zhao:2011hg,Li:2010dy,Guo:2011wr,Liu:2011wi,Liu:2009mga,
Castro:2010au,Correa:2010zg,Castro:2010uj,Chumbes:2010xg,Castro:2011pp}. 
In order to analyze resonances, we must compute the transmission coefficient ($T$), that gives a clear physical interpretation of what happens to a free wave interacting with the membrane. 
The idea of the existence of a resonant mode is that, for a given mass, the transmitted and reflected oscillatory modes are in phase inside the membrane. The transmission coefficient has a peak at 
that mass value, meaning that the amplitude of the wave-function has a maximum value at $z=0$ and the probability to find this KK mode inside the membrane is higher. This method has been used previously to analyze resonant modes of gravity, fermion and form fields \cite{Landim:2011ki,Landim:2011ts}.

The background considered here consists of a symmetric $Z_2$ thick domain wall interpolated between two BPS vacua. The scalar field is important because, in some cases, its behaviour is very similar to the gravitational field`s behaviour \cite{Kehagias:2000au}. The gauge vector field is an important ingredient of the standard model and, even being a not localizable field in theories with conformal symmetries, we may understand how its resonances appear due to the specific brane chosen. In this case, we consider the corrections due to a dilaton field coupled to the gauge field and others antisymmetrical tensors fields. These antisymmetric tensor fields arise quite naturally in string theory \cite{Polchinski:1998rq,Polchinski:1998rr} and supergravity \cite{VanNieuwenhuizen:1981ae} and play an important role in the dualization processes \cite{Smailagic:1999qw,Smailagic:2000hr}. In particular they appear in the $R-R$ sector of each of the type II string theories. These tensor fields couple naturally to higher-
dimensional extended objects, the $D-$branes, and are important for their stability. From a more mathematical point of view, they are related to the linking number of higher dimensional knots \cite{Oda:1989tq}. The rank of these antisymmetric tensors is defined by the dimension of the manifold\cite{Nakahara:2003nw}. Beyond this, these kind of fields play an important role in the solution of the moduli stabilization problem of string theory\cite{Kachru:2002he,Antoniadis:2004pp,Balasubramanian:2005zx}.

For the physics of extra dimensions is important to study higher rank tensor fields in membrane backgrounds. In this context, the antisymmetric tensor fields have already been considered in models of extra dimension. Generally speaking, the $q-$forms of highest rank do not have physical relevance because when the rank increases the number of gauge freedom increases as well. Such fact can be used to cancel the dynamics of the field in the brane\cite{Alencar:2010vk}. The spectrum mass of the two and three-form have been studied in Refs. \cite{Mukhopadhyaya:2004cc} and \cite{Mukhopadhyaya:2007jn} in the context of five dimensions with codimension one. The coupling between two and three-forms with the dilaton was also studied but in a different context \cite{DeRisi:2007dn,Mukhopadhyaya:2009gp,Alencar:2010mi}. An analysis 
of localization and computations of resonances due to $q$-forms have been previously made by the authors in \cite{Landim:2012zz}.

Here, in this piece of work we show results of particular importance to situations where the gravity backgrounds are smooth generalizations of the Randall-Sundrum model and procedures that are applied even when the Schr\"odinger potential is not known analytically \cite{Csaki:2000fc,Bazeia:2007nd}. More precisely, we continue the analysis made in \cite{Cvetic:2008gu} in order to understand the behaviour of scalar, gauge vector fields and more general $q$-forms in a model that can be solved analytically. Another goal is to make a comparison between the calculation obtained through the solution of the analytical method and the solutions obtained by the method of transmission coefficients. The importance of such a model with analytical solution resides in the fact that in the models with thick membranes their thickness is a constant parameter. When the thickness is parametrized, there is a way  to obtain the thin limit and the thick membrane limit. In this sense, the physics related now can be understood by 
means of a more general situation that can mimics several types of membranes. Through this comparison process we can establish the Transfer Matrix method, once an for all, as a tool to understand some physical aspects of extra dimensions.

The paper is organized as follows. In the first section we review the method of computation of transmission coefficients in order to make comparisons between different formalisms. In the second section we present the general procedure to solve the Schr\"odinger equation coming from models of extra dimensions and solve it for the scenario proposed. In the third section we quickly remember the gravitational results and make a generalization regarding solutions as plane waves far from the membrane.  The remaining sections are devoted to studies about the several bosonic fields, i.e., the antisymmetric gauge fields that usually appear in literature. They are the scalar field, gauge vector and tensor (Kalb-Ramond) fields, besides more general $q$-forms. Finally we present our conclusions and perspectives of work. 

\section{A Simple Example}

In this section we give a review of the Transfer Matrix method and present a model with known analytical solution. After that we compare results of both the analytical and numerical transmission coefficient computations. The analytical model discussed here is already known and can be found in \cite{LL}. 
It is important to say that all potentials considered in this manuscript are volcano like, see Fig. \ref{fig:multistep2}.

\FIGURE{
\centerline{\psfig{figure=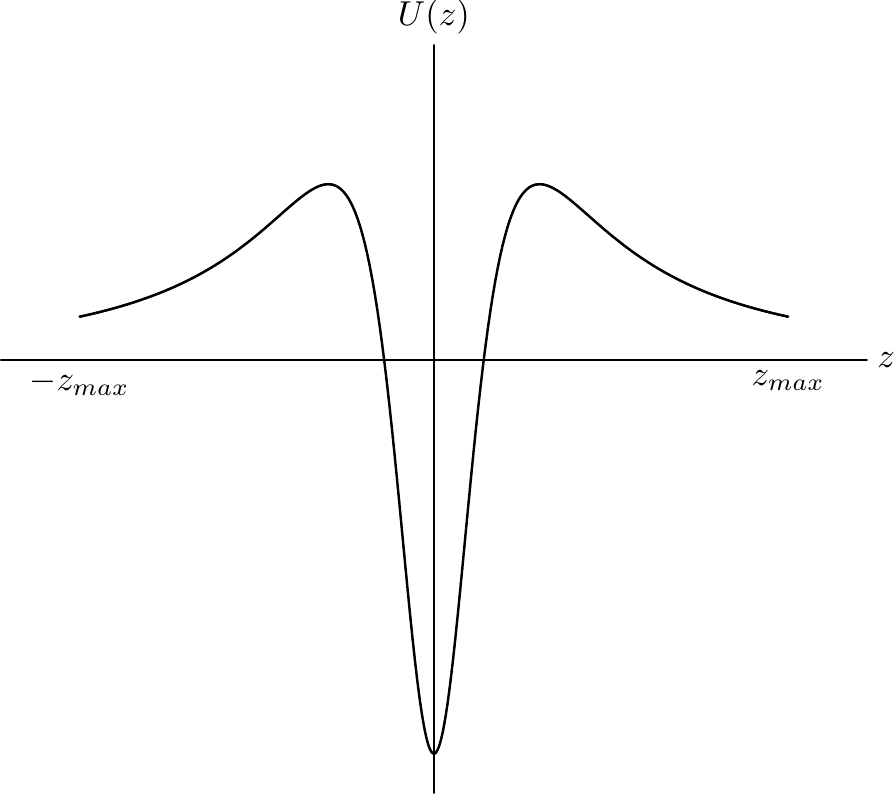,angle=0,height=5cm}}
\caption{General potential with parity symmetry with $\lim_{z \rightarrow \pm\infty}U(z)=0$. }\label{fig:multistep2}
}

The potential given in Fig.\ref{fig:multistep2} can be approximated by a series of potential barriers. In each region showed in the Fig.\ref{fig:multistep4}
\FIGURE{
\centerline{\psfig{figure=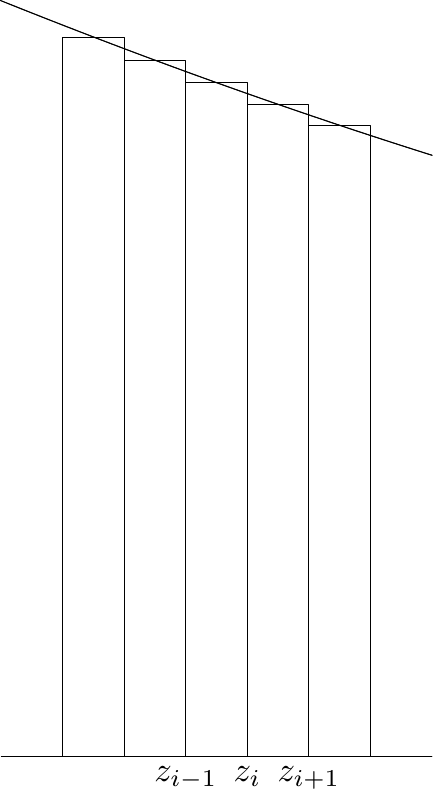,angle=0,height=6cm}}
\caption{The multistep regions.}\label{fig:multistep4}
}
the Schr\"odinger equation can be solved for the interval $z_{i-1}<z<z_i$, where the potential can be approximated by 
\begin{equation}
U(z)=U(\overline{z}_{i-1})=U_{i-1},\quad\overline{z}_{i-1}=(z_i+z_{i-1})/2.
\end{equation}

As the potential is null in the infinity the solution must be a plane wave. Then, as in Fig.\ref{fig:multistep5}, 
\FIGURE{
\centerline{\psfig{figure=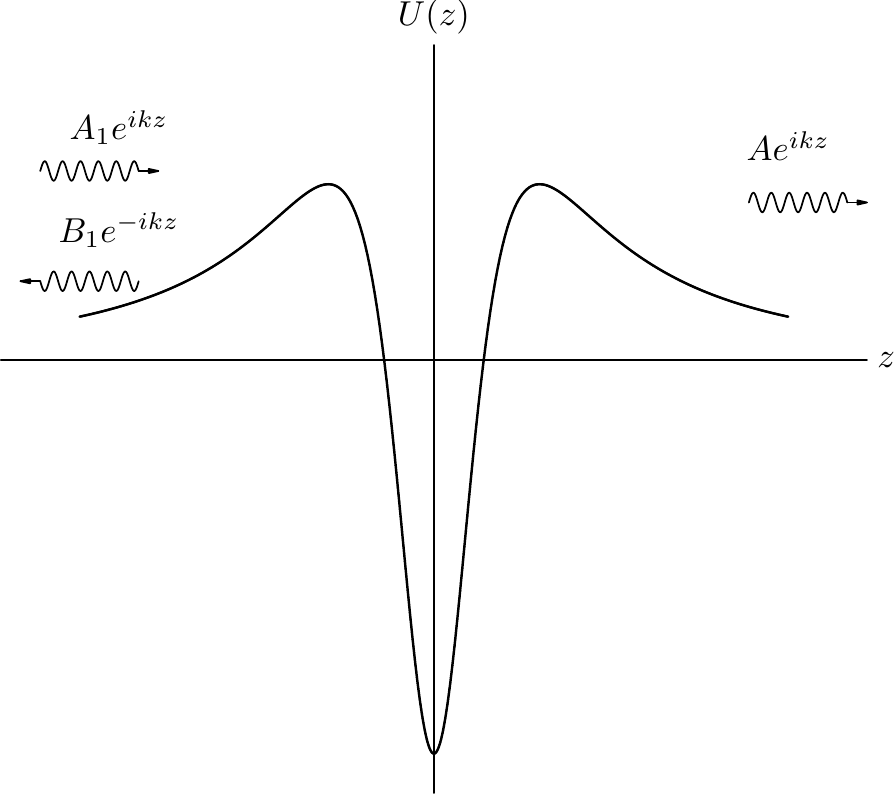,angle=0,height=6cm}}
\caption{The scattering of incident wave into reflected and transmitted waves.}\label{fig:multistep5}
}
we consider a plane wave colliding with the membrane. This solution choice is very important for the analytical solution considered in the next section. As the final goal here is to compare the analytical and the numerical results, we will consider plane waves as boundary 
conditions to our analytical solution. Therefore, the solution in each interval is
\begin{equation}
\psi_{i-1}(z)=A_{i-1}e^{ik_{i-1}z}+B_{i-1}e^{-ik_{i-1}z},\quad k_{i-i}=\sqrt{\lambda-U_{i-1}},
\end{equation}
and the continuity of the $\psi_{i-1}(z)$ and $\psi'_{i-1}(z)$ at $z=z_i$ gives us
\begin{equation}
\left(
\begin{array}{c}
 A_i\\ 
B_i
\end{array} \right)
=
M_i\left(
\begin{array}{c}
 A_{i-1}\\ 
B_{i-1}
\end{array} \right).
\end{equation}
In the above equation we have that
\begin{equation}
M_i=
\frac{1}{2k_i}\left(\begin{array}{cc}
 (k_i+k_{i-1})e^{-i(k_i-k_{i-1})z_i}& (k_i-k_{i-1})e^{-i(k_i+k_{i-1})z_i} \\ 
(k_i-k_{i-1})e^{i(k_i+k_{i-1})z_i} & (k_i+k_{i-1})e^{i(k_i-k_{i-1})z_i}
\end{array} \right)
\end{equation}
and performing this procedure iteratively we reach to
\begin{equation}
\left(
\begin{array}{c}
 A_N\\ 
B_N
\end{array} \right)
=
M\left(
\begin{array}{c}
 A_{0}\\ 
B_{0}
\end{array} \right),
\end{equation}
where,
\begin{equation}
M=M_NM_{N-1}\cdots M_{2}M_1.
\end{equation}
The transmission coefficient is therefore given by
\begin{equation}
T=1/|M_{22}|^2.
\end{equation}
In order to numerically obtain resonances we choose $z_{max}$ to 
satisfy $U(z_{max})\sim 10^{-4}$ and let $m^2$ runs from $U_{min}=U(z_{max})$ to $U_{max}$ (the maximum potential value). We divide $2z_{max}$ by $10^4$ or $10^5$ such that we have $10^4+1$ or $10^5+1$ transfer matrices.

Now we turn our attention to a model with analytical solution. Consider a Schr\"odinger equation with a potential given by \cite{LL}
\begin{equation}\label{landaupot}
 U(z)=\frac{U_0}{\cosh^2{\alpha z}},
\end{equation}
with profile as in Fig. \ref{fig:profile-landau}. 
\FIGURE{
\centerline{\psfig{figure=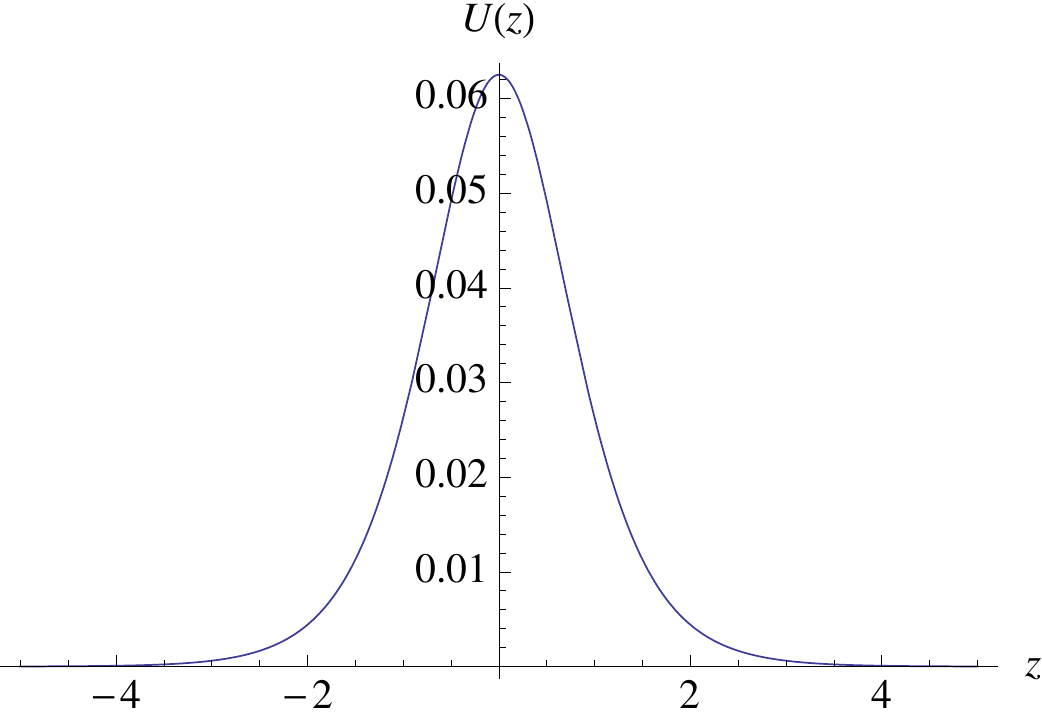,angle=0,height=5cm}}
\caption{The profile of the potential in Ref. \cite{LL} with $U_0=1/16$ and $\alpha=1$.}\label{fig:profile-landau}
}

The analytical solution to the wave function for the potential given by Eq. (\ref{landaupot}) is 
\begin{equation}
 \psi=(1-\xi^{2})^{-ik/2\alpha}F(-i\frac{k}{\alpha}-s,-i\frac{k}{\alpha}+s+1,-i\frac{k}{\alpha}+1,\frac{1}{2}(1-\xi)),
\end{equation}
where
\begin{eqnarray}
 \xi&=&\tanh \alpha z,\qquad k=\sqrt{2mE}/\hbar, \nonumber
\\s&=&\frac{1}{2}\left(-1+\sqrt{1-\frac{8mU_0}{\alpha^{2}\hbar^{2}}}\right).
\end{eqnarray}

As stressed in \cite{LL} this solution has the desired asymptotic plane wave form. After some calculations we can find the transmission coefficient:

\begin{eqnarray}
 T&=&\frac{\sinh^2(\pi k/\alpha)}{\sinh^2(\pi k/\alpha)+\cos^2[\frac{1}{2}\sqrt{1-8mU_0/\hbar^2\alpha^2}]},\quad 8mU_0/\hbar^2\alpha^2<1, \nonumber
\\
T&=&\frac{\sinh^2(\pi k/\alpha)}{\sinh^2(\pi k/\alpha)+\cosh^2[\frac{1}{2}\sqrt{8mU_0/\hbar^2\alpha^2-1}]},\quad 8mU_0/\hbar^2\alpha^2>1.
\end{eqnarray}

Now we can compare the results obtained by the analytical  and numerical  calculations. We use $\alpha=\hbar=m=1$ and $U_0=1/16$. Fig. \ref{fig:landau} shows: (a) The graphic for the analytical transmission coefficient $T_A$, (b) The graphic for the numerical transmission coefficient $T_N $ and (c) The graphic with the  ratio $T_N/T_A$. There is no distinction between both methods, the agreement is almost complete. 
\FIGURE{
\centerline{\psfig{figure=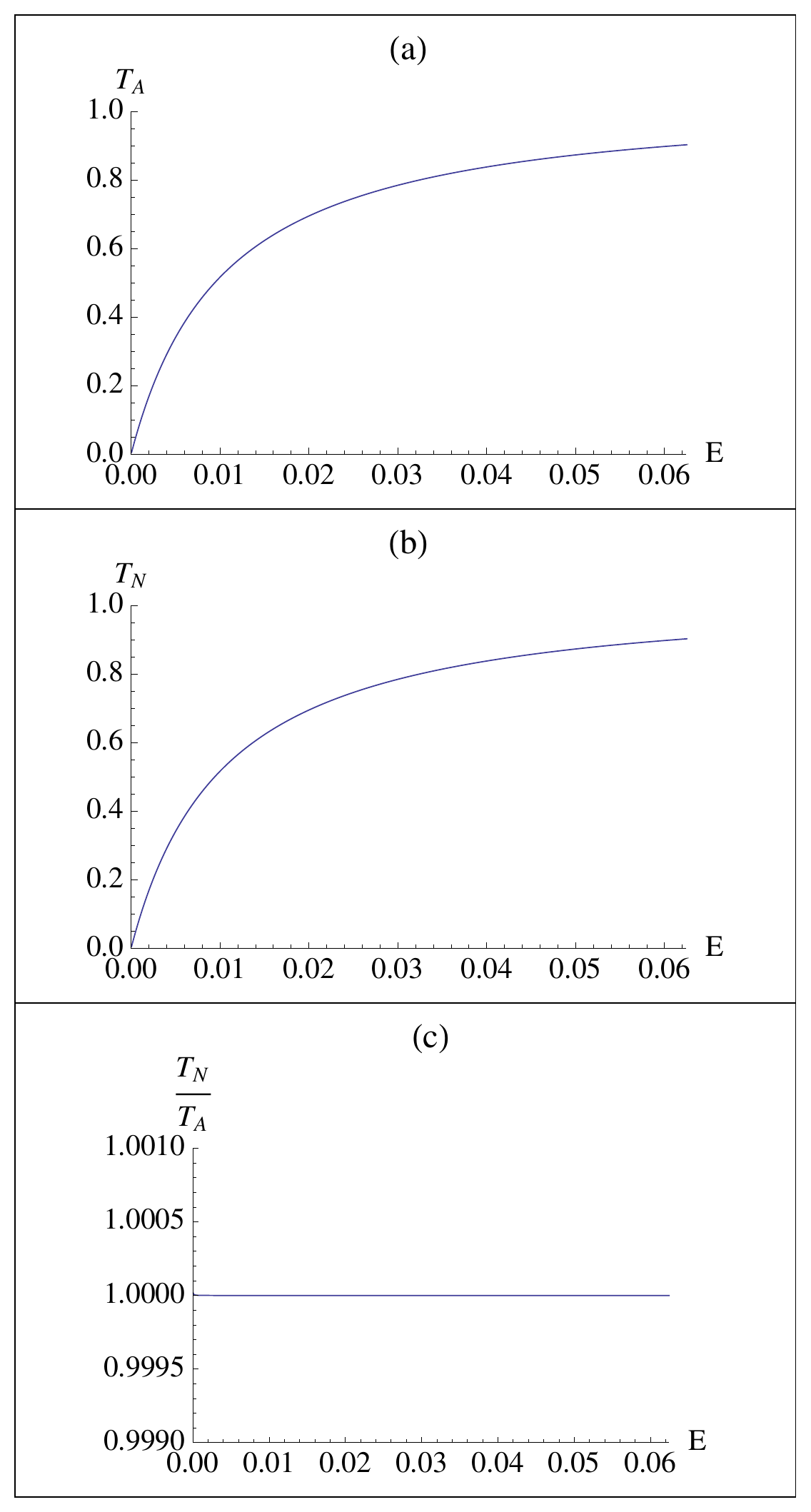,angle=0,height=9cm}}
\caption{Transmission coefficients plots in the Landau book example: (a) The analytical transmission coefficient $T_A$, (b) The numerical transmission coefficient $T_N$ and (c) The ratio $T_N/T_A$.}\label{fig:landau}
}

\section{The General Analytical Solution}

In this section we show the procedure used through all this work. In models describing interactions between fields and membranes, the equations of motion coming from the extra-dimension dependence 
plays an important role, and have a Schr\"odinger like final form. To arrive at that this result we provide a short review here (details can be found in \cite{Kehagias:2000au}). First of all we consider 
the line element of the background space as 
\be{line}
ds^2=e^{2A(y)}\eta_{\mu\nu}+e^{2B(y)}dy^2,
\ee
where $\eta_{\mu\nu}=\mbox{diag}(-1,1,1,1)$. This metric comes from the inclusion of the interaction produced by a dilaton field that, in a specific model, gives rise to localization of gauge fields.

The Einstein`s equations give the relation $B=A/4$ between the functions $A(y)$ and $B(y)$ \cite{Kehagias:2000au}. These equations are obtained from the usual superpotential method in models 
containing kink like defects. It is important to cite that the dilaton field have been already been 
treated in the Randall-Sundrum scenario \cite{Alencar:2010mi}. As we are going to consider models with and without the dilaton ($B=0$), we introduced a constant $b$ to take account of this and we get $B(y)=(1-b)A(y)$. 
The value of $b$ with the dilaton is $3/4$ and $1$ without the dilaton. 
Using this metric background, in general, the equations we need to work have the following form
\be{beq}
\left(-\frac{d^{2}}{dy^{2}}+P'(y)\frac{d}{dy}+V(y)\right)\psi(y)=m^{2}
Q(y)\psi(y),
\ee
where $P(y)=\gamma A(y)$, $Q(y)=e^{-2bA(y)}$ and $V(y)=0$ for all fields except gravity, in which $V(y)=2A''(y)-2(1+b)A'(y)^2$. Here, $\gamma$ is a constant that depends on the field.
We can transform \equ{beq} in a Schr\"odinger like equation through the transformations

\begin{equation}
\frac{dz}{dy}=f(y),\quad\psi(y)=\Omega(y)\overline{\psi}(z),\label{transformation}
\end{equation}
with
\begin{equation}
f(y)=\sqrt{Q(y)},\quad\Omega(y)=\exp(P(y)/2)Q(y)^{-1/4},
\end{equation}
and 
\begin{equation}
\bar{U}(z)=V(y)/f^{2}+\left(P'(y)\Omega'(y)-\Omega''(y)\right)/\Omega f^{2},
\end{equation}
where the prime means the derivative in $y$. We emphasize here, that the
above expression is useful when $dz/dy=f(y)$
do not have an analytical solution. When this solution is known it
is better to express the potential in terms of derivatives in the $z$
coordinate:

\begin{equation}
\bar{U}(z)=\bar{V}(z)/\bar{f}^{2}(z)+\frac{\bar{P}'(z)\bar{\Omega}'(z)-\bar{
\Omega}''(z)}{\bar{\Omega}(z)}
-\frac{\bar{\Omega}'(z)}{\bar{\Omega}(z)}\frac{\bar{f}'(z)}{\bar{f}(z)},
\end{equation}
where $f(y)=\bar{f}(z)$. When the above steps are performed we get a Schr\"odinger like equation
\begin{equation}
 \{-\frac{d^2}{dz^{2}}+\bar{U}\}\bar{\psi}(z)=m^{2}\bar{\psi}(z),
\end{equation}
with potential  $\bar{U}(z)$ given by

\begin{equation}
 \bar{U}(z)=c\bar{A}^{''}(z)+c^{2}(\bar{A}'(z))^{2},
\end{equation}
where $c=-(\gamma+b)/2$ for all fields and $c=3/2$ for gravity.

An analytical solution has been obtained previously for the gravity case in \cite{Cvetic:2008gu} but without the dilaton coupling. We use the same method to solve the case with the dilaton coupling and for other fields. The relation between the $z$ and $y$ coordinates in all cases is given by $\frac{dz}{dy}=e^{|y|}=e^{-b\bar{A}(z)}$, for
$|z|>d/2$,where  $d$ is a constant. With this, we obtain

\be{aa}
A(y)=\bar{A}(z)=-\frac{1}{b}\ln (|z|+\beta).
\ee
The above expression determines the Schr\"odinger equation potential in the region $|z|>\frac{d}{2}$ as

\begin{equation}
\bar{U}(z) =  \frac{a}{(|z|+\beta)^{2}},
\end{equation}
with $a= \frac{c}{b}+\frac{c^{2}}{b^{2}}$. Since we can have  resonance only for positive potential, we restrict $a$ to be positive, i. e, $c/b>0$ or $c/b<-1$. For the region $|z|\leq d/2$, we choose

\begin{equation}
A(z)=\frac{1}{c}\ln\cos(\sqrt{V_{0}}|z|),
\end{equation}
such that

\begin{equation}
\bar{U}(z)=-V_{0}, ~|z|\leq d/2.
\end{equation}

Continuity of the metrics and its derivative at $z=\pm d/2$
give us the relations

\begin{eqnarray*}
(\frac{d}{2}+\beta) & =\cos(\sqrt{V_{0}}\frac{d}{2})^{-b/c},\\
(\frac{d}{2}+\beta) & ^{-1}=\frac{b}{c}\sqrt{V_{0}}\tan(\sqrt{V_{0}}\frac{d}{2}).
\end{eqnarray*}
In order to obtain $V_0$, $\beta$ and $d$  we will introduce the parameter $x=d\sqrt{V_{0}}/2$ as in \cite{Cvetic:2008gu}.

For the region $|z|>d/2$, the Schr\"odinger equation can be solved if we consider
$\bar{\psi}(z)=\sqrt{|z|+\beta}\bar{\phi}(z)$. With this transformation we obtain

\begin{equation}
\sqrt{|z|+\beta}\bar{\phi}''+\frac{1}{\sqrt{|z|+\beta}}\bar{\phi}'+\sqrt{
|z|+\beta}\left[m^{2}-\frac{1}{4(|z|+\beta)^{2}}-\frac{a}{(|z|+\beta)^{2}}\right]\bar{\phi}=0.
\end{equation}

Multiplying now by $(|z|+\beta)^{3/2}$ and defining $u=m(|z|+\beta)$
we arrive at

\begin{equation}
u^{2}\bar{\phi}''+u\bar{\phi}'+\left[u^{2}-\nu^{2}\right]\bar{\phi}=0,
\end{equation}
where the prime means a $u$ derivative and $\nu^{2}=(\frac{1}{2}+\frac{c}{b})^2$. This is a Bessel equation of order $\nu$. Here we are interested in solutions behaving like plane waves when $z\to\pm\infty$. Solutions with this properties are given by

\begin{equation}
\begin{cases}
H_{\nu}^{(1)}(u) & =J_{\nu}(u)+iY_{\nu}(u),\\
H_{\nu}^{(2)}(u) & =J_{\nu}(u)-iY_{\nu}(u),
\end{cases}
\end{equation}
with asymptotic behavior

\begin{equation}
\begin{cases}
\lim_{z\to\infty}H_{\nu}^{(1)}(u) & =\sqrt{\frac{2}{\pi
u}}e^{i(z-\frac{\pi}{4}-\frac{\nu\pi}{2})}\\
\lim_{z\to\infty}H_{\nu}^{(2)}(u) & =\sqrt{\frac{2}{\pi
u}}e^{-i(z-\frac{\pi}{4}-\frac{\nu\pi}{2})}.
\end{cases}
\end{equation}

The choice of asymptotic plane waves is related to the fact that we
want to compare this solution with the transfer matrix method, where
a plane wave colliding with the brane is used. It is important to stress here that this is different from
the solution found in \cite{Cvetic:2008gu}, where an exact solution was used
but no plane wave was considered. Using this consideration, the solution is given by

\begin{equation}
\psi(z)=\sqrt{\frac{u}{m}}\left(AH_{\nu}^{(1)}(u)+BH_{\nu}^{(2)}(u)\right),
\end{equation}
with the desired behavior.

\begin{equation}
\lim_{z\to\infty}\psi(z)=\sqrt{\frac{2}{{m\pi}}}\left(Ae^{imz}+Be^{-imz}\right).
\end{equation}
In the region $|z|\leq d/2$, we have the solution

\begin{equation}
\psi_{I}(z)=ae^{iKz}+be^{-iKz},
\end{equation}
with $K=\sqrt{m^{2}+V_{0}}$. The subscript $I$ means that the
solution is in the central region and all the above constants must
be fixed by the boundary conditions. As we are interested in calculate
resonances we must consider a plane wave coming from $-\infty$. This
plane wave will collide with the membrane and will generate a reflected
and a transmitted wave. Therefore for $z<-d/2$ we must have a linear
combination of waves moving to the left and to the right. For $z>d/2$ we must
have only one wave moving to the right. We define the solution in both
regions by $\psi_{L}(z)$ and $\psi_{R}(z)$ respectively. In order to analyze resonances
we fix the coefficient of the incoming wave equal to one. Therefore we choose 

\begin{eqnarray*}
\psi_{L}(z) & = & \sqrt{\frac{u}{m}}\left(H_{\nu}^{(2)}(u)+B_{1}H_{\nu}^{(1)}(u)\right),~z<0,\\
\psi_{R}(z) & = & \sqrt{\frac{u}{m}}A_{2}H_{\nu}^{(1)}(u),~z>0.
\end{eqnarray*}

Defining $E(z)=\sqrt{\frac{u}{m}}H_{\nu}^{(2)}(u),F(z)=\sqrt{\frac{u}{m}}H_{\nu}^{(1)}(u)$ for $z<0$ and taking the continuity of the wave function and its derivative at $z=\pm d/2$, we obtain

\begin{equation}
\begin{pmatrix}E(-d/2) & F(-d/2)\\
E'(-d/2) & F'(-d/2)
\end{pmatrix}\begin{pmatrix}1\\
B_{1}
\end{pmatrix}=\begin{pmatrix}e^{-iK\frac{d}{2}} & e^{iK\frac{d}{2}}\\
iKe^{-iK\frac{d}{2}} & -iKe^{iK\frac{d}{2}}
\end{pmatrix}\begin{pmatrix}a\\
b
\end{pmatrix},
\end{equation}
and  

\begin{equation}
\begin{pmatrix}F(-d/2) & 0\\
-F'(-d/2) & 0
\end{pmatrix}\begin{pmatrix}A_{2}\\
0
\end{pmatrix}=\begin{pmatrix}e^{iK\frac{d}{2}} & e^{-iK\frac{d}{2}}\\
iKe^{iK\frac{d}{2}} & -iKe^{-iK\frac{d}{2}}
\end{pmatrix}\begin{pmatrix}a\\
b
\end{pmatrix}.
\end{equation}
After some algebra we finally obtain
\be{amplitude}
A_2=\frac{-4iK}{\pi(2KFF'\cos Kd+(F'^2-F^2K^2)\sin Kd)},
\ee
with the transmission coefficient  given by

\be{analictt}
T=|A_2|^2=\frac{16K^2}{\pi^2|2KFF'\cos Kd+(F'^2-F^2K^2)\sin Kd|^2}.
\ee

That is the final expression we use as starting point to make comparison between analytical and numerical methods. It is important to mention the number of iterations made using the method applied here. For calculations involving the dilaton contribution we have 
used $10^5$ matrices, 
$2\times10^5$ matrices for the gravity case with $x=1.5$, and $10^5$ matrices with $x=1.0$ for the scalar and gauge field cases. We have made another 
calculation of $2\times10^5$ matrices for the parameter $x=0.15$  again for the vector gauge field resonances ($x$ is the parameter describing the thickness 
of the membrane). For the tensor fields we computed  $10^5$ matrices for $x=1.5$ and $x=1.0$. For the case without the dilaton contribution, we 
have computed $3\time10 ^5$ matrices for $x=1.5$ and $10^5$ for $x=1.0$ for the scalar field resonances. In the case of vector fields we made $10^5$ iterations 
with $x=1.0$ plus another computation of $10^5$ matrices for $x=0.15$. It is important to mention that we have found, for the case without dilaton, the same 
results for resonances in the gravity and scalar field cases. The results of these calculations will be discussed below.

\section{The Gravity Case}

As mentioned before the gravity case without dilaton has been considered previously in \cite{Cvetic:2008gu}. We will reconsider it here for completeness. The potential for this case
is given by

\begin{equation}
\bar{U}_1(z)=\frac{3}{4}e^{2A_1(y)}\left(2A{''}(y)+5(A_1{'}(y))^{2}\right).
\end{equation}
The transformation to the $z$ coordinate is obtained using $\frac{dz}{dy}=e^{-\bar{A}_1(z)}$ to get

\begin{equation}
\bar{U}_1(z)=\frac{3}{2}\bar{A}_1{''}(z)+\frac{9}{4}(\bar{A}_1^{'}(z))^{2}\label{
potential},
\end{equation}
what is equivalent to choose $b=1$ and $c=3/2$ in the previous section. From this we can
obtain the solution taking $\nu=2$. 

The case with the dilaton coupling has the potential

\begin{equation}
\bar{U}_2(z)=\frac{3}{2}e^{3A_2(y)/2}\left(A_2{''}(y)+\frac{9}{4}(A_2'(y))^{2}\right).
\end{equation}
Where using the transformation $\frac{dz}{dy}=e^{-\frac{3}{4}\bar{A}_2(z)}$ we
obtain

\begin{equation}
\bar{U}_{2}(z)=\frac{3}{2}\bar{A}_{2}{''}(z)+\frac{9}{4}(\bar{A}_{2}'(z))^{2},
\end{equation}
what is equivalent to choose $b=3/4$ and $c=3/2$. The solution
is given by taking $\nu=5/2$. Using now the expression for the transmission
coefficient we obtain the results with dilaton  in Fig. \ref{fig:gravity-x10-comdilaton}. In the figure
we can see the appearance of one peak of resonance for the parameters $x=1.0$ and $\lambda\sqrt{3M^3}=20$. These peaks of resonance are very dependent on the parameter $x$, when we take  $x=1.5$ and $\lambda\sqrt{3M^3}=20$ the transmission coefficient for gravity with dilaton presents seven peaks of resonances shown in Fig. \ref{fig:gravity-x15-comdilaton}.

\FIGURE{
\centerline{\psfig{figure=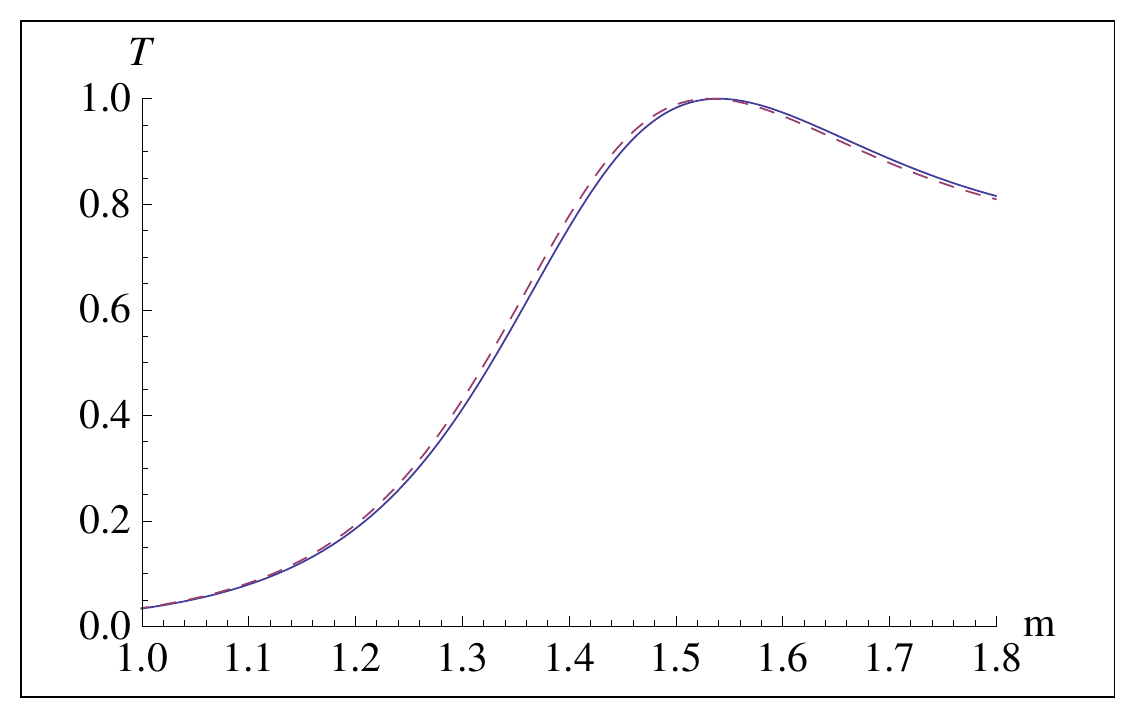,angle=0,height=6cm}}
\caption{The transmission coefficient for gravity with dilaton for $x=1.0$ and $\lambda\sqrt{3M^3}=20$. The solid line is the analytical calculation and the dashed line is the numerical one.}\label{fig:gravity-x10-comdilaton}
}

\FIGURE{
\centerline{\psfig{figure=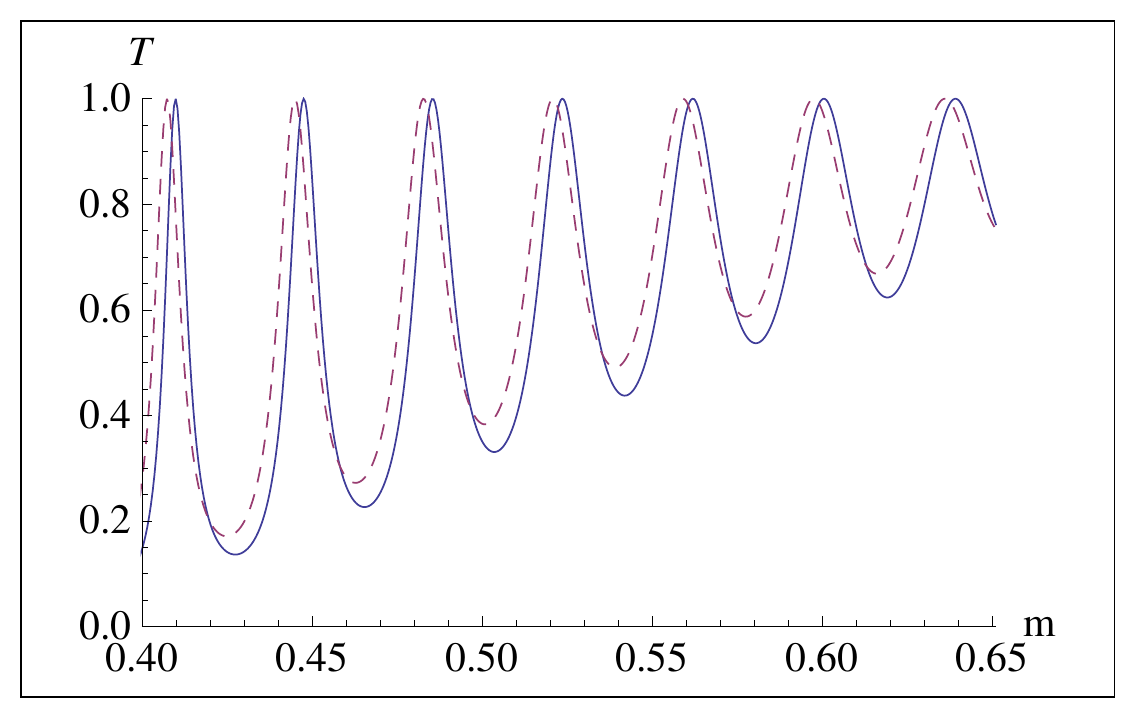,angle=0,height=6cm}}
\caption{The transmission coefficient for gravity with dilaton for $x=1.5$ and $\lambda\sqrt{3M^3}=20$. The solid line is the analytical calculation and the dashed line is the numerical one.}\label{fig:gravity-x15-comdilaton}
}

\section{The Scalar Field Case}

Now we analyze the scalar field resonances by the analytical model proposed. As in the previous
section we must obtain the potential in the $z$ coordinate for the
cases with and without the dilaton. These are given by

\begin{eqnarray}
 &&\bar{U}_{1}(z)=e^{3A_1(y)/2}\left(-(\frac{\alpha}{2}+\frac{3}{8})A_1(y){''}+(\frac{
\alpha}{4}-\frac{9}{64})(A_1'(y))^{2}\right),\\
&&\bar{U}_{2}(z)=e^{2A_2(y)}\left(\frac{3}{2}A_2{''}(y)+\frac{15}{4}(A_2'(y))^{2}\right),
\end{eqnarray}
where $\alpha=-15/4-\lambda\sqrt{3M^{3}}$.  Performing now the
transformations $\frac{dz}{dy}=e^{-3\bar{A}_1(z)/4}$ and $\frac{dz}{dy}=e^{-\bar{A}_2(z)}$
respectively, we obtain
\begin{eqnarray}
\bar{U}_{1}(z)&=&\left(-(\frac{\alpha}{2}+\frac{3}{8})\bar{A}_1{''}(z)+(\frac{\alpha}{2
}+\frac{3}{8})^{2}(\bar{A}_1'(z))^{2}\right) \nonumber\\
&=&\left((\frac{3}{2}+\frac{\lambda\sqrt{3M^{3}}}{2})\bar{A}_1{''}(z)+(\frac{3}{2}+\frac{\lambda\sqrt{3M^{3}}}{2})^{2}(\bar{A}_1'(z))^{2}\right),
\end{eqnarray}
and
\begin{eqnarray}
 \bar{U}_{2}(z)&=&\left(\frac{3}{2}\bar{A}_2{''}(z)+\frac{9}{4}(A_2'(z))^{2}\right).
\end{eqnarray}

Therefore the solutions to both cases can be found by using
\begin{eqnarray*}
b_{1} & = & \frac{3}{4},\quad c_{1}=(\frac{3}{2}+\frac{\lambda\sqrt{3M^{3}}}{2}),\\
b_{2} & = & 1,\quad c_{2}=\frac{3}{2},
\end{eqnarray*}
and we get
\begin{eqnarray*}
\nu_{1}^{2} & = &
(\frac{5}{2}+\frac{2\lambda\sqrt{3M^{3}}}{3})^2
\\
\nu_{2} & = & 2.
\end{eqnarray*}

We now compare these results with those obtained from the numerical method of transmission coefficients. We show in Fig. \ref{fig:scalar-x10-semdilaton} the transmission coefficient for analytical (lined) and 
numerical (dashed)  without dilaton with $x=1.0$, showing one peak of resonance. In Fig. \ref{fig:scalar-x10-comdilaton} with dilaton field we still have one peak of resonance for the parameters $x=1.0$ 
and $\lambda\sqrt{3M^3}=20$. In Fig. \ref{fig:scalar-x15-semdilaton} we have the Transmission coefficient for scalar without dilaton for $x=1.5$ and $\lambda\sqrt{3M^3}=20$ and, in Fig. \ref{fig:scalar-x15-comdilaton} 
showing eight peaks of resonances (the first one near $m=0$, can be made more visible if we just give a zoom in that region), the Transmission coefficient for scalar with dilaton for $x=1.5$ and $\lambda\sqrt{3M^3}=20$ 
showing three peaks.

\FIGURE{
\centerline{\psfig{figure=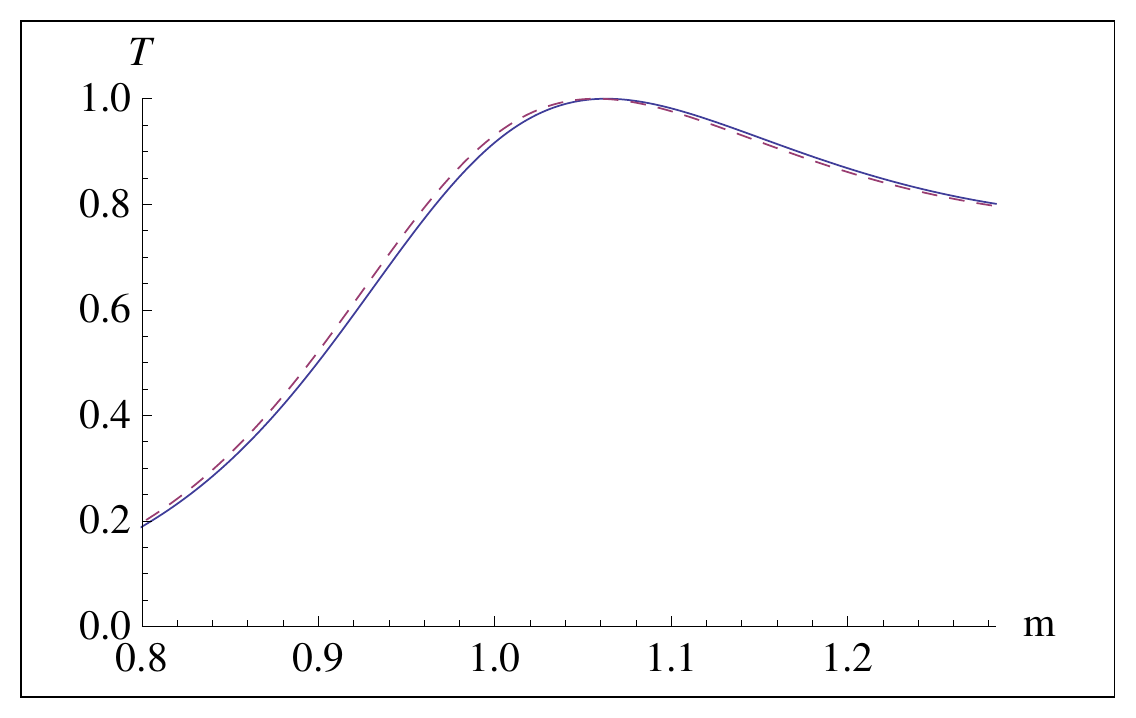,angle=0,height=6cm}}
\caption{Transmission coefficient for scalar without dilaton for $x=1.0$. Lined is analytic and dashed is numeric. }\label{fig:scalar-x10-semdilaton}
}

\FIGURE{
\centerline{\psfig{figure=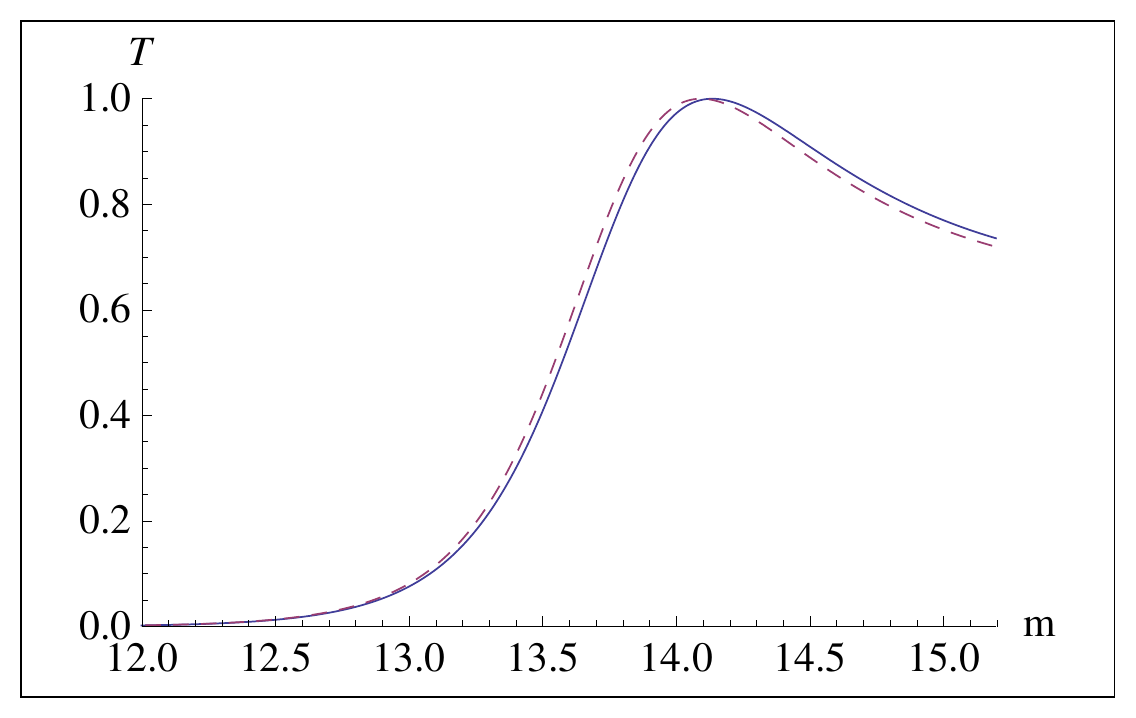,angle=0,height=6cm}}
\caption{Transmission coefficient for scalar with dilaton for $x=1.0$ and $\lambda\sqrt{3M^3}=20$. Lined is analytic and dashed is numeric.}\label{fig:scalar-x10-comdilaton}
}

\FIGURE{
\centerline{\psfig{figure=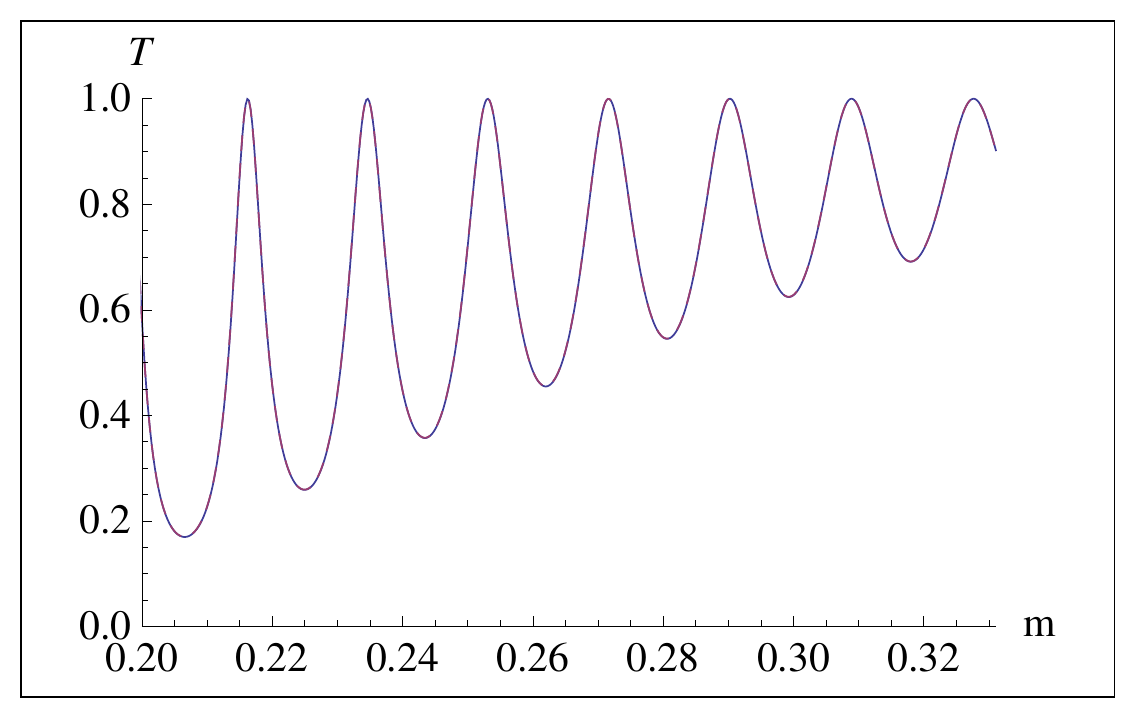,angle=0,height=6cm}}
\caption{Transmission coefficient for scalar without dilaton for $x=1.5$ and $\lambda\sqrt{3M^3}=20$. Lined is analytic and dashed is numeric.}\label{fig:scalar-x15-semdilaton}
}

\FIGURE{
\centerline{\psfig{figure=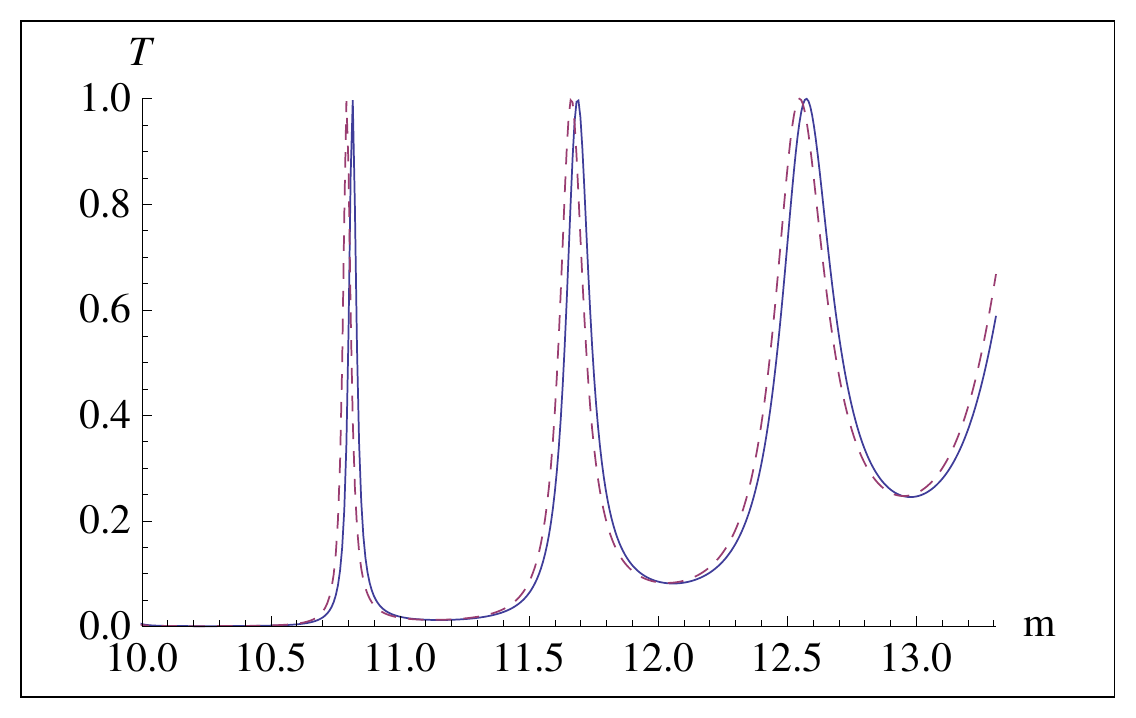,angle=0,height=6cm}}
\caption{Transmission coefficient for scalar with dilaton for $x=1.5$ and $\lambda\sqrt{3M^3}=20$. Lined is analytic and dashed is numeric.}\label{fig:scalar-x15-comdilaton}
}

\section{The Gauge Field Case}

Now we turn our attention to analyze the gauge field resonances through the analytical context. Again we need to obtain the potential
in the $z$ coordinate for the cases with and without the dilaton. These potentials are given by
\begin{eqnarray}
&&\bar{U}_{1}(z)=e^{3A_1(y)/2}\left(-(\frac{\alpha}{2}+\frac{3}{8})A_1{''}(y)+(\frac{
\alpha}{4}-\frac{9}{64})(A_1'(y))^{2}\right),\\
&&\bar{U}_{2}(y)=e^{2A(y)}\left(\frac{1}{2}A{''}(y)+\frac{3}{4}(A'(y))^{2}\right),
\end{eqnarray}
where $\alpha=-7/4-\lambda\sqrt{3M^{3}}$. Performing now the transformations
$\frac{dz}{dy}=e^{-3\bar{A}_1(z)/4}$ and $\frac{dz}{dy}=e^{-\bar{A}_2(z)}$ respectively,
we obtain

\begin{eqnarray}
\bar{U}_{1}(z)&=&\left(-(\frac{\alpha}{2}+\frac{3}{8})\bar{A}_1{''}(z)+(\frac{\alpha}{2
}+\frac{3}{8})^{2}(\bar{A}_1'(z))^{2}\right)\nonumber
\\
&=&\left((\frac{1}{2}+\frac{\lambda\sqrt{3M^{3}}}{2})\bar{A}_1{''}(z)+(\frac{1}{2}+\frac{\lambda\sqrt{3M^{3}}}{2})^{2}(\bar{A}_1'(z))^{2}\right),
\end{eqnarray}
and
\begin{eqnarray}
 \bar{U}_{2}(z)=\left(\frac{1}{2}\bar{A}_2{
''}(z)+\frac{1}{4}(\bar{A}_2'(z))^{2}\right). 
\end{eqnarray}

Therefore the solutions to both cases can be found by using 
\begin{eqnarray*}
b_{1} & = & \frac{3}{4},\quad c_{1}=(\frac{1}{2}+\frac{\lambda\sqrt{3M^{3}}}{2}),\\
b_{2} & = & 1,\quad c_{2}=\frac{1}{2},
\end{eqnarray*}
what gives for the solutions with and without the dilaton 

\begin{eqnarray*}
\nu_{1}^{2} & = &(\frac{7}{6}+\frac{2\lambda\sqrt{3M^{3}}}{3})^2,
\\
\nu_{2} & = & 1.
\end{eqnarray*}

We show in Fig. \ref{fig:vector-x10-comdilaton} and in Fig. \ref{fig:vector-x10-semdilaton}  the transmission coefficient for analytical (lined) and numerical (dashed) with and  without dilaton with $x=1.0$ showing 
one peak of resonance, and in Fig. \ref{fig:vector-x15-comdilaton} we show the three peaks of the transmission coefficient for vector field with dilaton for $x=1.5$ and $\lambda\sqrt{3M^3}=20$.

\FIGURE{
\centerline{\psfig{figure=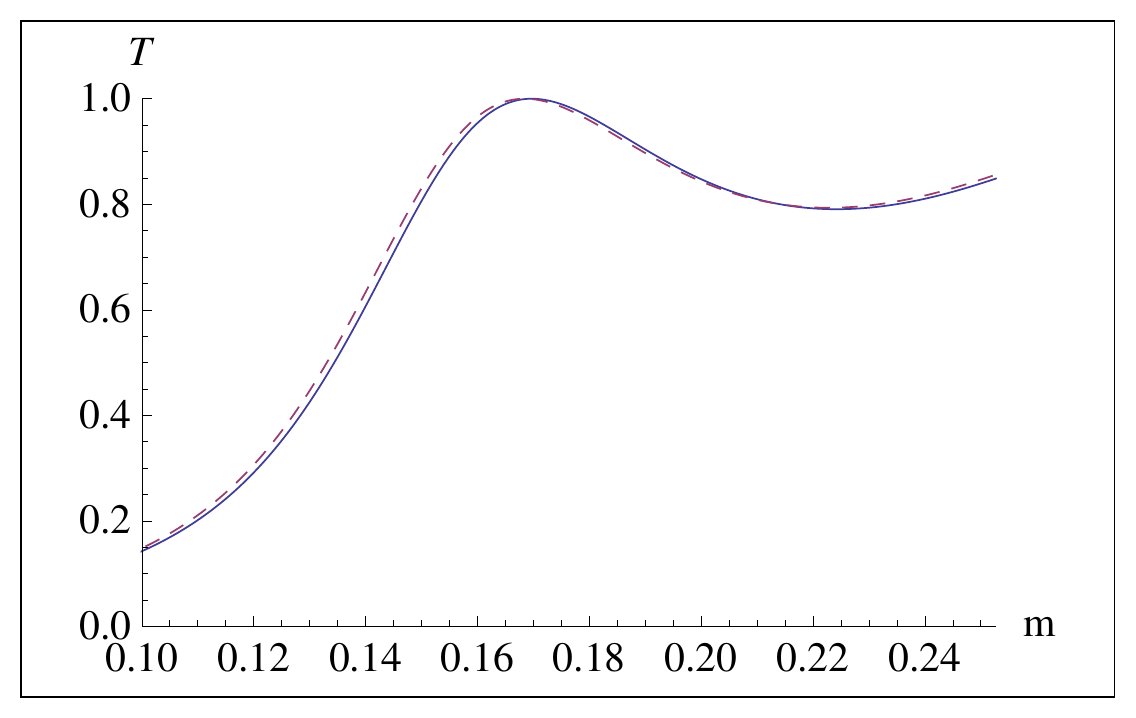,angle=0,height=6cm}}
\caption{Transmission coefficient for vector field without dilaton for $x=1.0$. Lined is analytic and dashed is numeric.}\label{fig:vector-x10-semdilaton}
}

\FIGURE{
\centerline{\psfig{figure=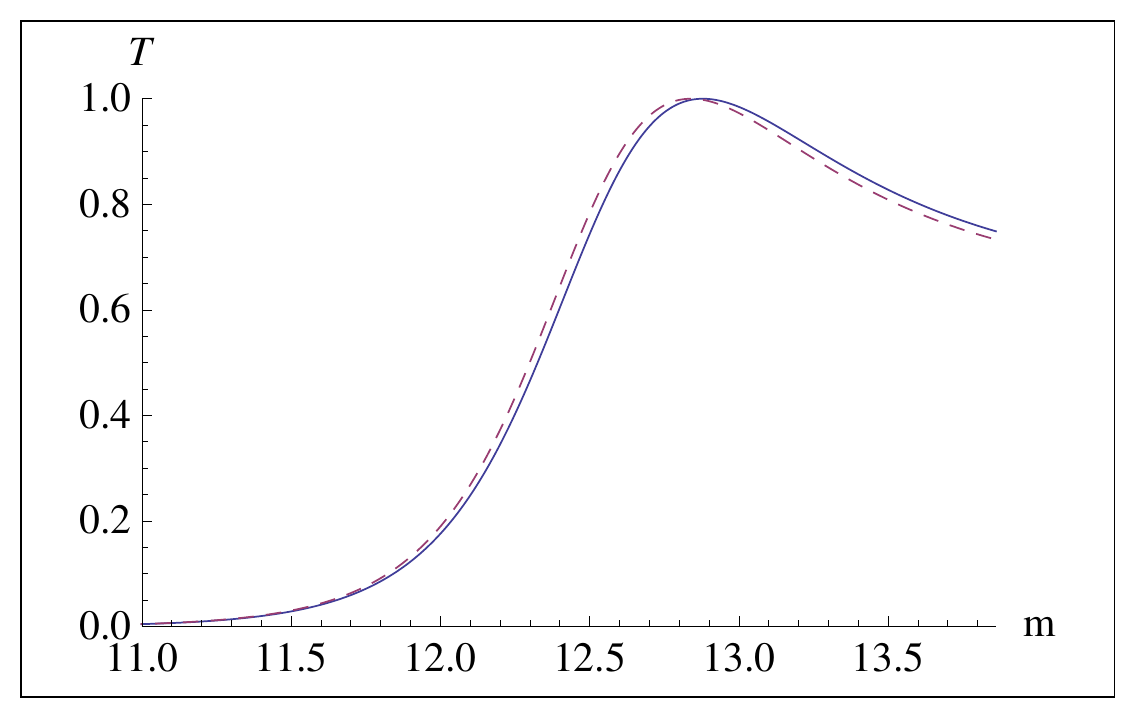,angle=0,height=6cm}}
\caption{Transmission coefficient for vector field with dilaton for $x=1.0$ and $\lambda\sqrt{3M^3}=20$. Lined is analytic and dashed is numeric.}\label{fig:vector-x10-comdilaton}
}

\FIGURE{
\centerline{\psfig{figure=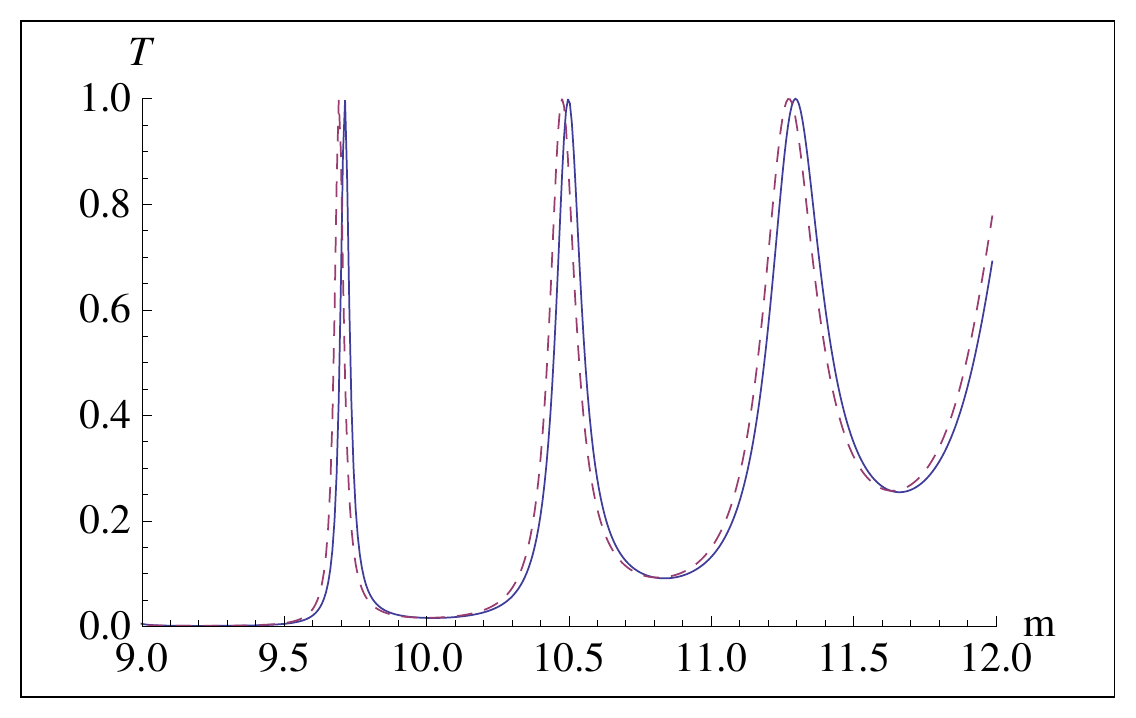,angle=0,height=6cm}}
\caption{Transmission coefficient for vector field with dilaton for $x=1.5$ and $\lambda\sqrt{3M^3}=20$. Lined is analytic and dashed is numeric.}\label{fig:vector-x15-comdilaton}
}
\section{The Kalb-Ramond Case}

The last bosonic field to be analyzed in five dimensions is the Kalb-Ramond field. Now we analyze this field resonances in the same way as in the last sections. The potential in the $z$ coordinate for the cases 
with and without the dilaton are given by
\begin{eqnarray}
&&\bar{U}_{1}(y)=e^{3A_1(y)/2}\left(-(\frac{\alpha}{2}+\frac{3}{8})A_1{''}(y)+(\frac{
\alpha}{4}-\frac{9}{64})(A_1'(y))^{2}\right),\\
&&\bar{U}_{2}(y)=e^{2A_2(y)}\left(\frac{1}{2}A_2{''}(y)+\frac{3}{4}(A_2'(y))^{2}\right),
\end{eqnarray}
where $\alpha=1/4-\lambda\sqrt{3M^{3}}$.  Performing now the transformations
$\frac{dz}{dy}=e^{-3\bar{A}_1(z)/4}$ and $\frac{dz}{dy}=e^{-\bar{A}_2(z)}$ respectively,
we obtain
\begin{eqnarray}
\bar{U}_{1}(z)&=&\left(-(\frac{\alpha}{2}+\frac{3}{8})\bar{A}_1{''}(z)+(\frac{\alpha}{2
}+\frac{3}{8})^{2}(\bar{A}_1'(z))^{2}\right)\nonumber
\\
&=&\left((-\frac{1}{2}+\frac{\lambda\sqrt{3M^{3}}}{2})\bar{A}_1{''}(z)+(-\frac{1}{2}+\frac{\lambda\sqrt{3M^{3}}}{2})^{2}(\bar{A}_1'(z))^{2}\right),
\end{eqnarray}
and
\begin{eqnarray}
 \bar{U}_{2}(z)=\left(-\frac{1}{2}\bar{A}_2{
''}(z)+\frac{1}{4}(\bar{A}_2'(z))^{2}\right). 
\end{eqnarray}

Therefore the solutions to both cases can be found by using 
\begin{eqnarray*}
b_{1} & = & \frac{3}{4},\quad c_{1}=(-\frac{1}{2}+\frac{\lambda\sqrt{3M^{3}}}{2}),\\
b_{2} & = & 1,\quad c_{2}=-\frac{1}{2},
\end{eqnarray*}
which gives for the solutions with and without the dilaton 

\begin{eqnarray*}
\nu_{1}^{2} & = &(-\frac{1}{6}+\frac{2\lambda\sqrt{3M^{3}}}{3})^2
\\
\nu_{2} & = & 0
\end{eqnarray*}

The case without the dilaton has $a=-1/4$ and therefore will not be considered because it gives a negative potential. 

We show in Fig. \ref{fig:tensor-x10-comdilaton} the transmission coefficient for analytical (lined) and numerical (dashed)  with dilaton with $x=1.0$ and in Fig. \ref{fig:tensor-x15-comdilaton} with $x=1.5$. In both cases we have $\lambda\sqrt{3M^3}=20$. In Fig. \ref{fig:tensor-x10-comdilaton} we see just one peak of resonance around $m=11.5$ and, in Fig. \ref{fig:tensor-x15-comdilaton} we have got five peaks.

\FIGURE{
\centerline{\psfig{figure=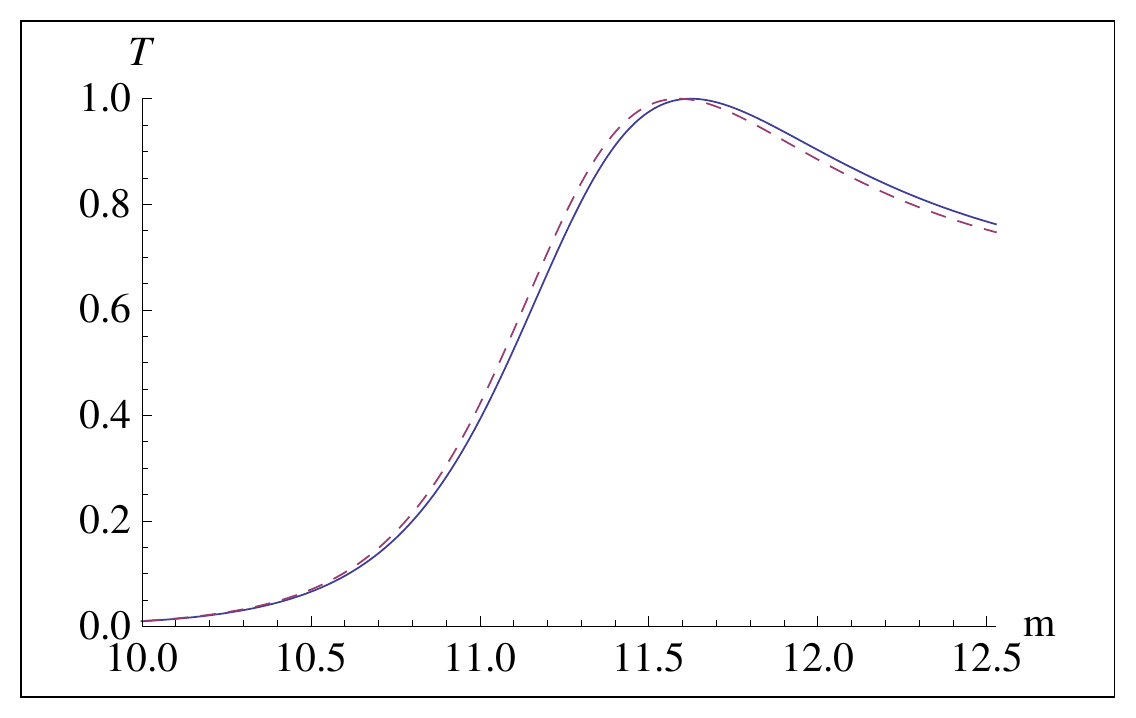,angle=0,height=6cm}}
\caption{Transmission coefficient for tensor field with dilaton for $x=1.0$ and $\lambda\sqrt{3M^3}=20$. Lined is analytic and dashed is numeric.}\label{fig:tensor-x10-comdilaton}
}

\FIGURE{
\centerline{\psfig{figure=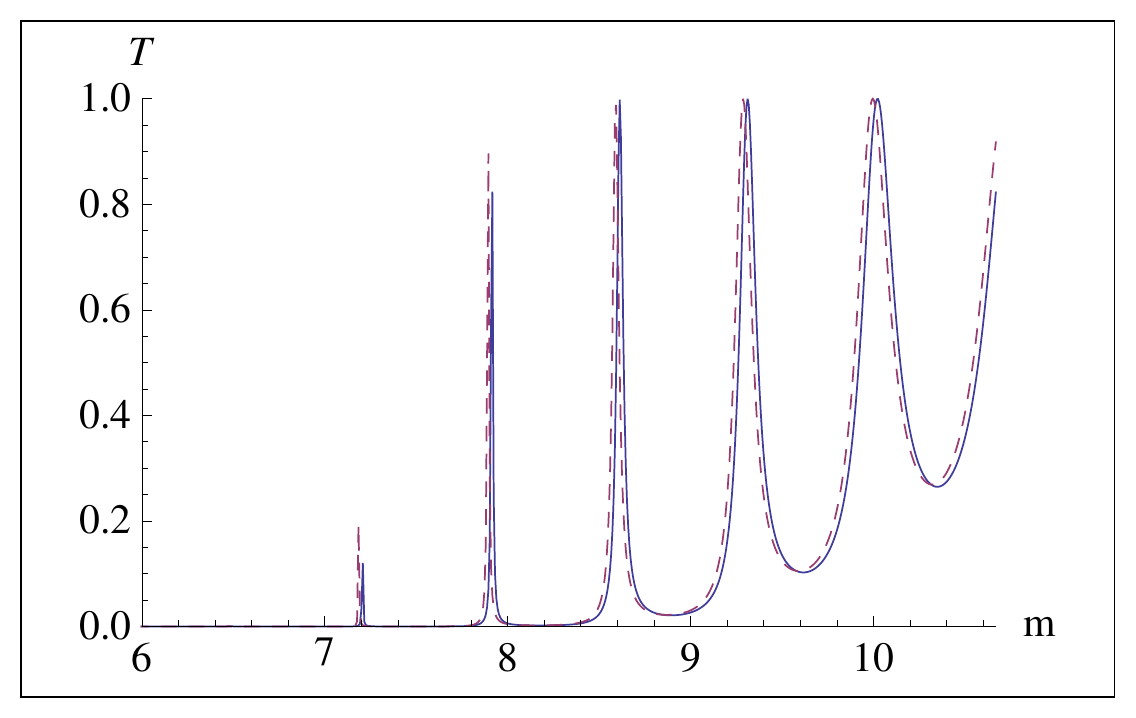,angle=0,height=6cm}}
\caption{Transmission coefficient for tensor field with dilaton for $x=1.5$ and $\lambda\sqrt{3M^3}=20$. Lined is analytic and dashed is numeric.}\label{fig:tensor-x15-comdilaton}
}
\section{The $q-$form Case}

The results of the last three cases can be summarized and generalized to a $q$-form in a $p$-brane, where $p=D-2$. In a recent article the present authors have studied the resonances of theses fields 
numerically in a different background\cite{Landim:2012zz}. There the potential in the $z$ coordinate for the case with the dilaton is given by
\begin{equation}
\bar{U}_{1}(z)=e^{3A_1(y)/2}\left(-(\frac{\alpha}{2}+\frac{3}{8})A_1''(y)+(\frac{
\alpha}{4}-\frac{9}{64})(A_1'(y))^{2}\right).
\end{equation}
In this equation $\alpha=(8q-4p-3)/4-\lambda\sqrt{3M^3}$. Performing the transformations
$\frac{dz}{dy}=e^{-3\bar{A}_1(z)/4}$ we obtain

\begin{eqnarray}
\bar{U}_{1}(z)&=&\left(-(\frac{\alpha}{2}+\frac{3}{8})\bar{A}_1{''}(z)+(\frac{\alpha}{2
}+\frac{3}{8})^{2}(\bar{A}_1'(z))^{2}\right).
\end{eqnarray}
For the case without the dilaton, we have obtained the potential
\begin{equation}
\bar{U}_2(z)=e^{2A_2(y)}\left((q-\frac{p}{2})(q-\frac{p}{2})(A_2'(y))^2-(q-\frac{p}{2})A''_2(y)\right),
\end{equation}
and using the transformation $\frac{dz}{dy}=e^{-\bar{A}_2(z)}$ we obtain
\begin{eqnarray}
 \bar{U}_{2}(z)=\left((\frac{p}{2}-q)\bar{A}_2
''(z)+(\frac{p}{2}-q)^{2}(A_2'(z))^{2}\right). 
\end{eqnarray}

Therefore the solutions for both cases can be found using 
\begin{eqnarray*}
b_{1} & = & \frac{3}{4},\quad c_{1}=(\frac{p}{2}-q+\frac{\lambda\sqrt{3M^3}}{2}),\\
b_{2} & = & 1,\quad c_{2}=(\frac{p}{2}-q),
\end{eqnarray*}
which gives the solutions for the cases with and without the dilaton 

\begin{eqnarray*}
\nu_{1}^{2} & = &(\frac{2\alpha}{3})^2
\\
\nu_{2}^{2} & = & (\frac{1+p}{2}-q)^2.
\end{eqnarray*}

The above formulas can be verified for the cases $p=3$ with $q=0,1,2$ and agree with the last sections. For the case with the dilaton we show results for the parameter $\alpha$ only. The condition for positivity of the  potential is $\alpha<-3/4$ or $\alpha>3/4$. For the case without the dilaton we only have the parameter $(p/2-q+1/2)$. For this case the condition for the positivity of the potential is given by $q<p/2$ or $q>p/2+1$ and therefore we understand why, for the two form the potential is negative in $p=3$. It is important to note that for $p>3$ higher forms can be analyzed. We show in Fig. \ref{fig:fx-semdil-c5-profile} the potential profile for a $q$-form field without dilaton with $p/2-q=5$.

For the $q$-form case the transmission coefficient is plotted in Fig. \ref{fig:tx10-comdil-alpha-1} for $x=1.0$ and  $\alpha=-1$ and in Fig. \ref{fig:tx10-comdil-alpha-2} for the same $x$  and $\alpha=-2$ considering the dilaton. For the case without the dilaton the transmission coefficient is shown in Fig. \ref{fig:tx10-semdil-c5} for $p/2-q=5$ and in Fig. \ref{fig:tx10-semdil-c7}  for $p/2-q=7$ , $x=1.0$ in both cases. One can see one peak of resonances in all figures except for Fig. \ref{fig:tx10-comdil-alpha-1} that has two peaks. For completeness, we show the transmission coefficient  in Fig. \ref{fig:tx15-semdil-c5} and Fig. \ref{fig:tx15-semdil-c7} without dilaton for $p/2-q=5$ and $p/2-q=7$. They show six and eight peaks of resonances respectively.

\FIGURE{
\centerline{\psfig{figure=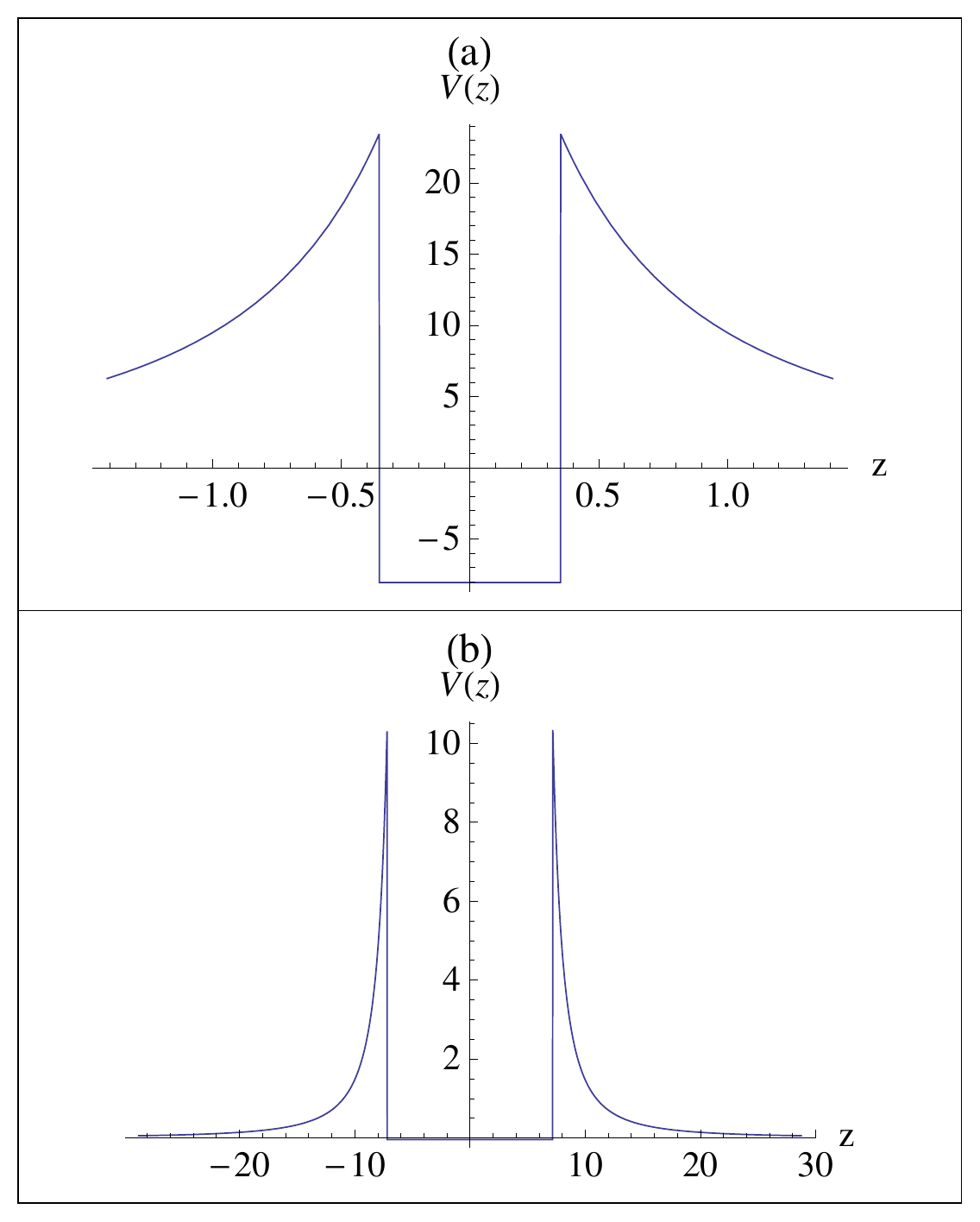,angle=0,height=8cm}}
\caption{The potential profile for $q-$form field without dilaton with $p/2-q=5$. (a) $x=1.0$ and (b) $x=1.5$.}\label{fig:fx-semdil-c5-profile}
}

\FIGURE{
\centerline{\psfig{figure=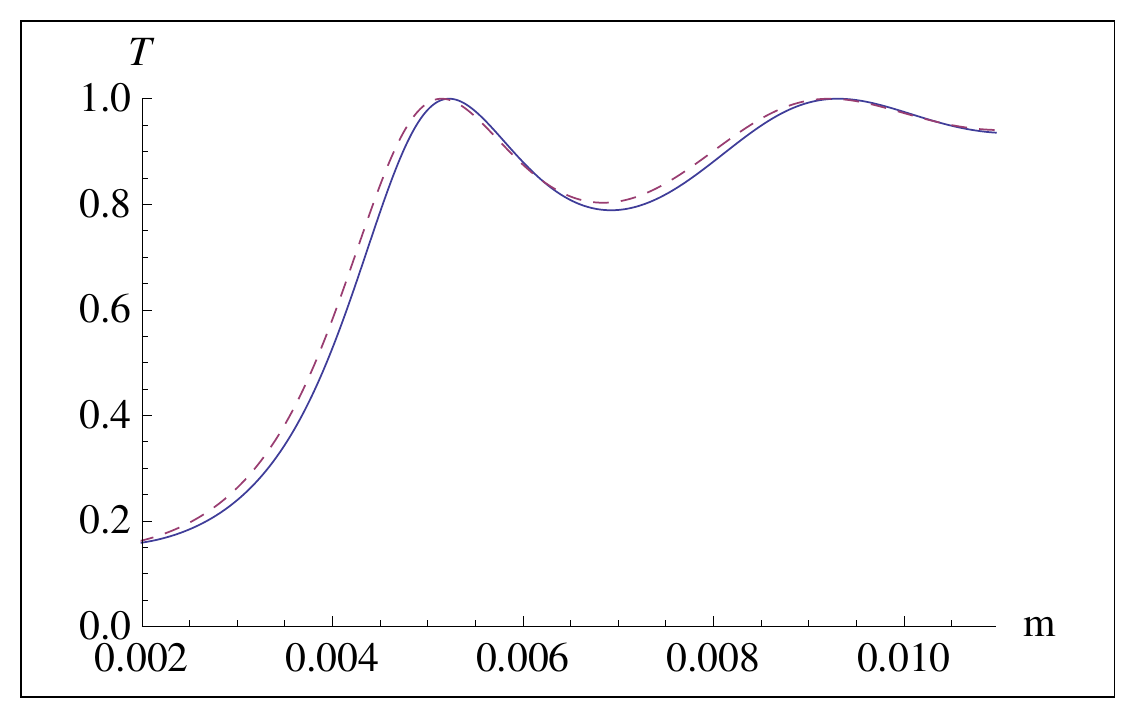,angle=0,height=5cm}}
\caption{Transmission coefficient for $q-$form field with dilaton for $x=1.0$ and $\alpha=-1$. The solid line is the analytical solution, the dashed line is the numerical solution.}\label{fig:tx10-comdil-alpha-1}
}

\FIGURE{
\centerline{\psfig{figure=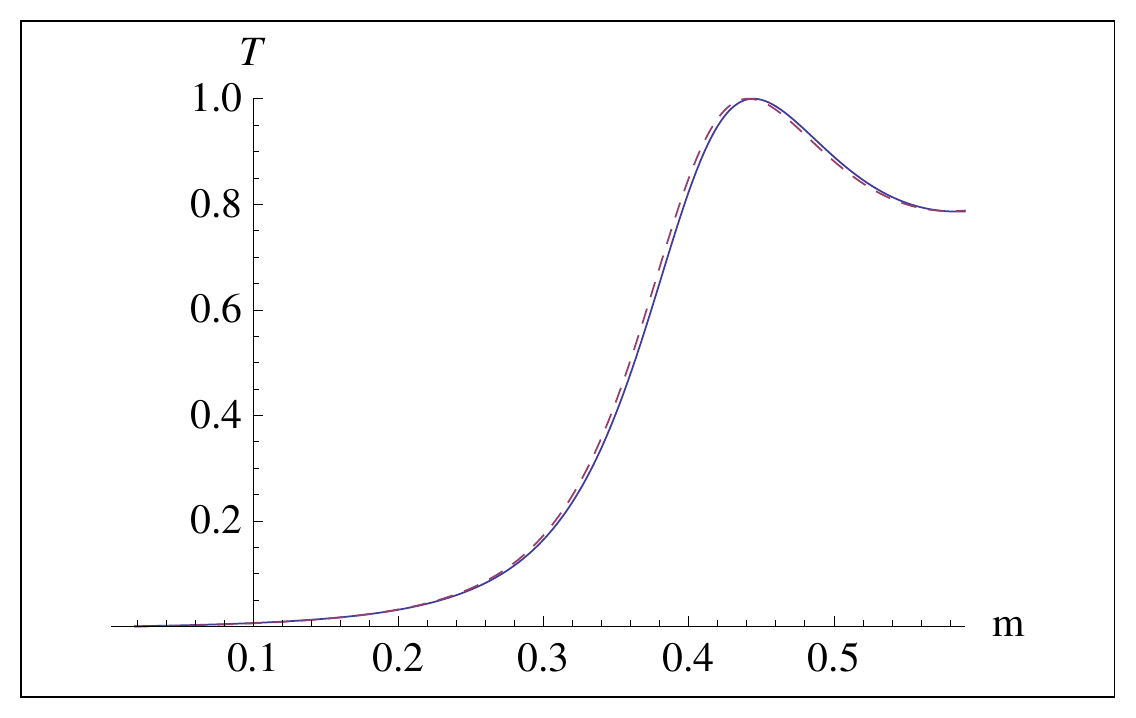,angle=0,height=5cm}}
\caption{Transmission coefficient for $q-$form field with dilaton for $x=1.0$ and $\alpha=-2$. The solid line is the analytical solution, the dashed line is the numerical solution  .}\label{fig:tx10-comdil-alpha-2}
}

\FIGURE{
\centerline{\psfig{figure= 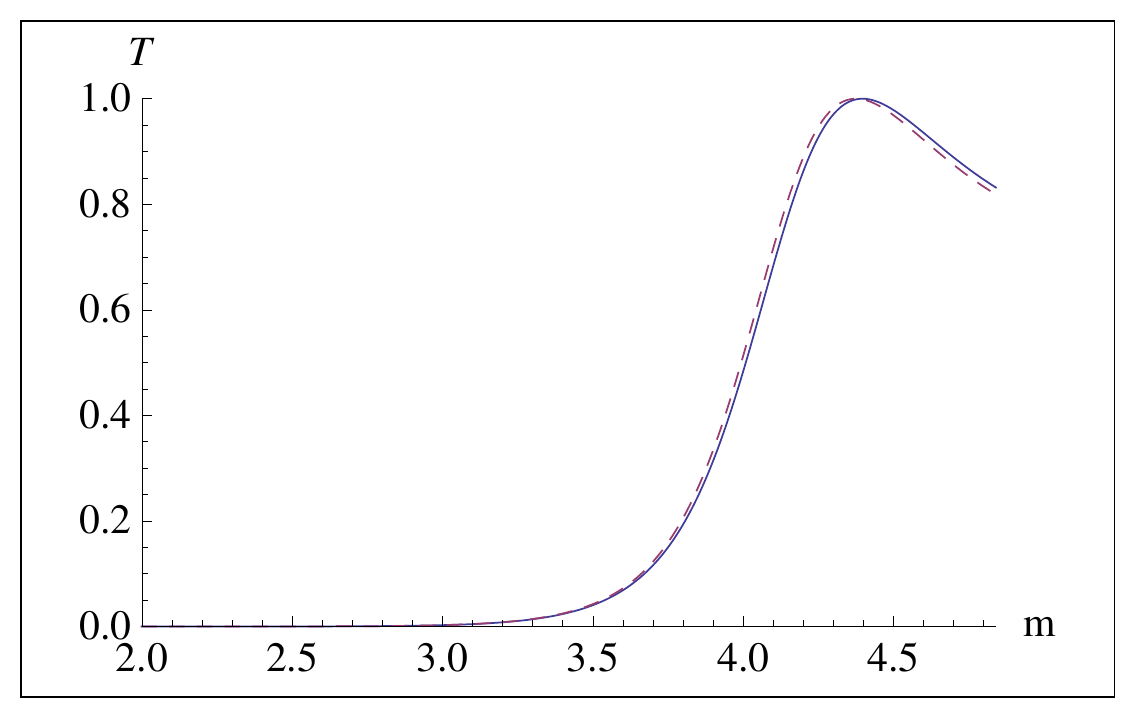,angle=0,height=5cm}}
\caption{Transmission coefficient for $q-$form field without dilaton for $x=1.0$ and $p/2-q=5$. The solid line is the analytical solution, the dashed line is the numerical solution .}\label{fig:tx10-semdil-c5}
}

\FIGURE{
\centerline{\psfig{figure=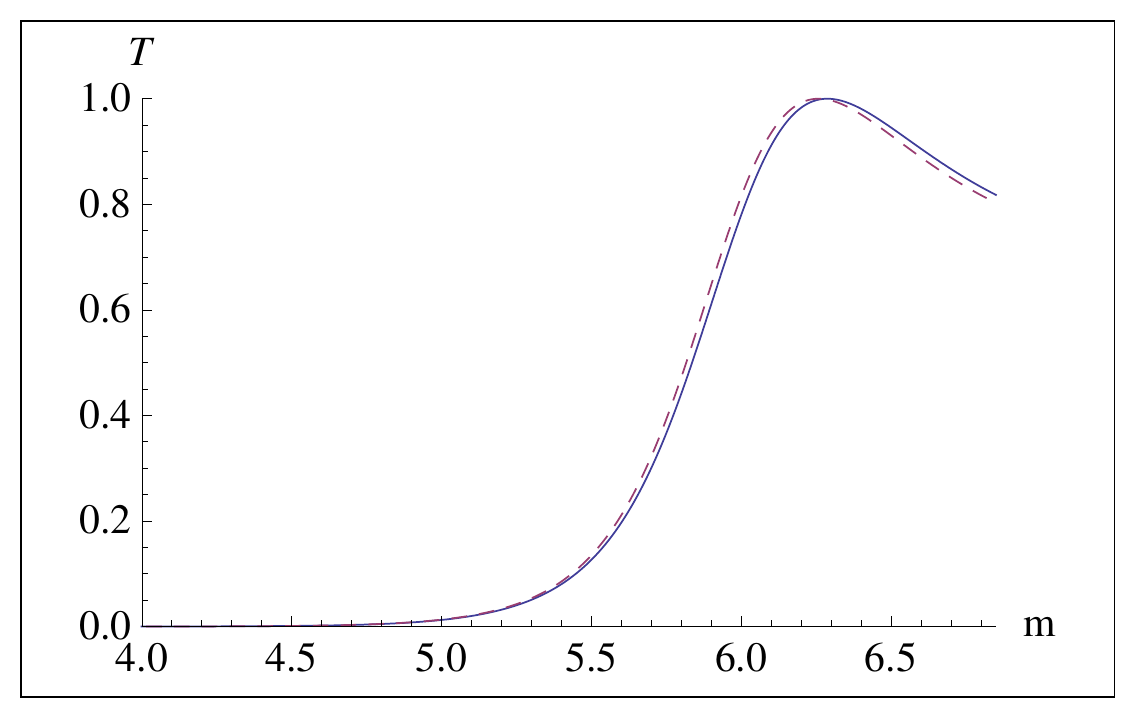,angle=0,height=5cm}}
\caption{Transmission coefficient for $q-$form field without dilaton for $x=1.0$ and $p/2-q=7$. The solid line is the analytical solution, the dashed line is the numerical solution .}\label{fig:tx10-semdil-c7}
}

\FIGURE{
\centerline{\psfig{figure= 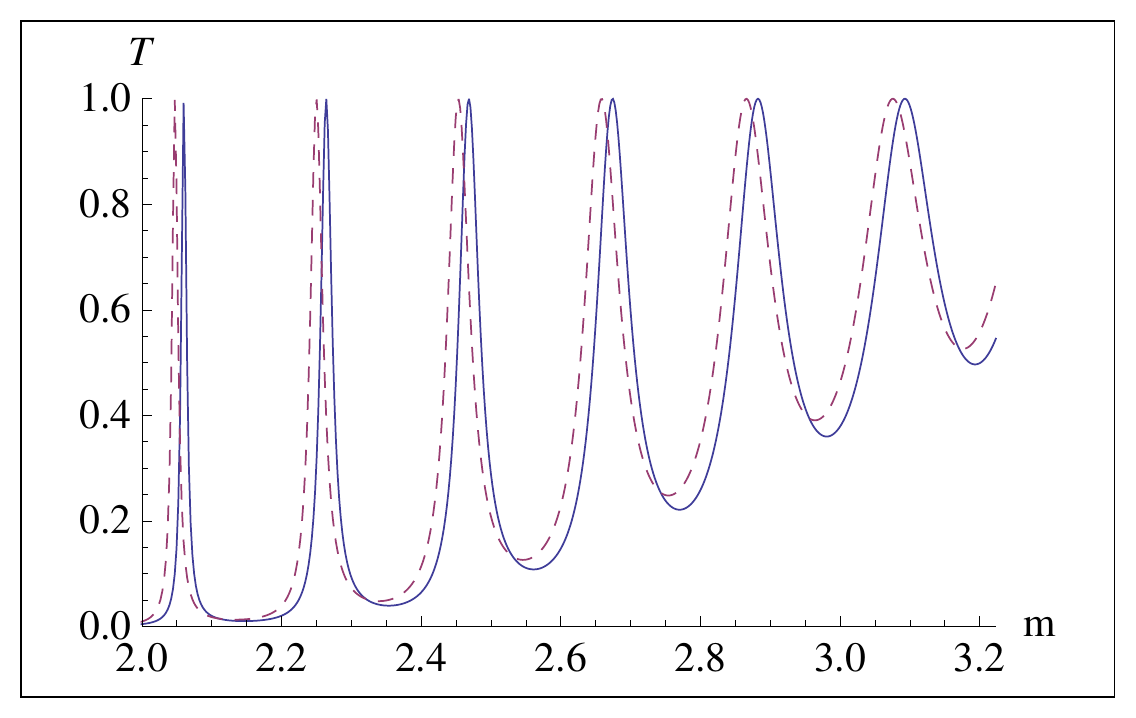,angle=0,height=5cm}}
\caption{Transmission coefficient for $q-$form field without dilaton for $x=1.5$ and $p/2-q=5$. The solid line is the analytical solution, the dashed line is the numerical solution .}\label{fig:tx15-semdil-c5}
}

\FIGURE{
\centerline{\psfig{figure=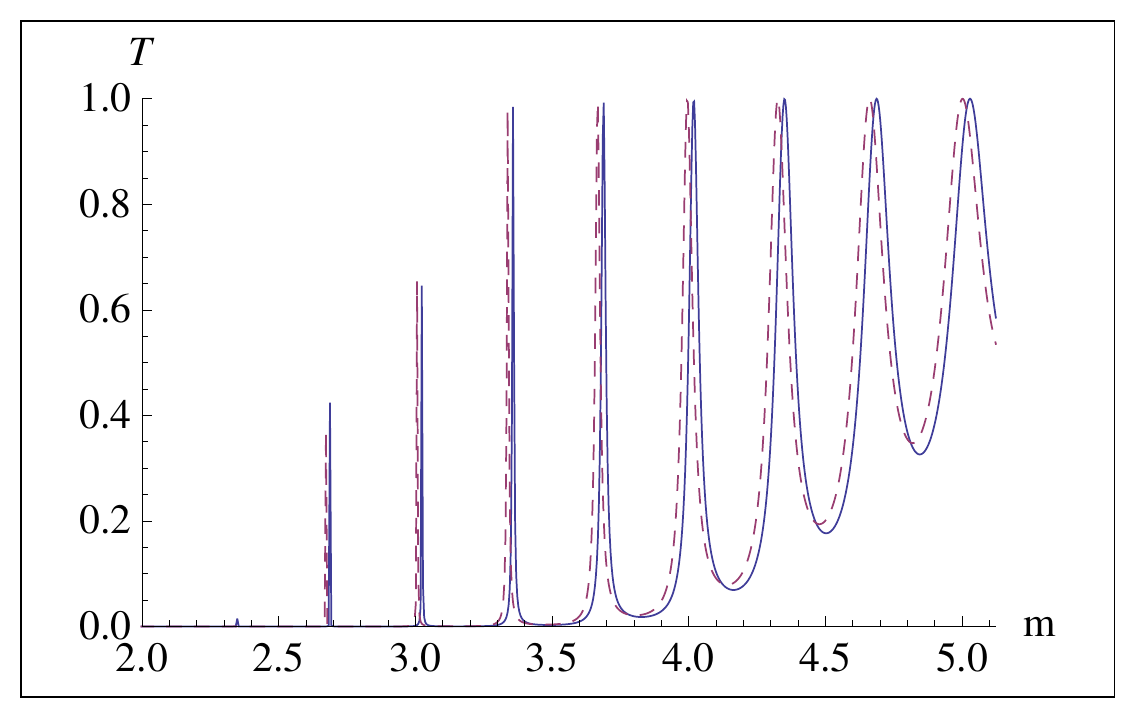,angle=0,height=5cm}}
\caption{Transmission coefficient for $q-$form field without dilaton for $x=1.5$ and $p/2-q=7$. The solid line is the analytical solution, the dashed line is the numerical solution .}\label{fig:tx15-semdil-c7}
}

\section{Conclusions and Perspectives}

In this work we have analyzed the transfer matrix method in the light of models with analytical solution.
The background considered consist of a symmetric $Z_2$ thick domain wall interpolating between two BPS vacua. It has been shown previously that this background allows for an analytical 
solution of the gravity field modes \cite{Cvetic:2008gu}. By following the same lines of reasoning, as a first result we have found exact solutions to all the bosonic fields, namely, the scalar, the 
vector and the KR field. More than this, our solution to the Gravity case is slightly different of ref. \cite{Cvetic:2008gu}. The reason is that in that work they considered the wave function as real. 
Here we need a complex wave function since we consider plane waves arriving from the infinity and colliding with the membrane. 

The first case used to test the numerical method was that of Ref. \cite{LL}. For this well known case the difference in the transmission coefficient values is of the order of $10^{-17}$. For the other 
fields namely: the scalar field, the vector field , the Kb and $q-$form fields the agreement was impressively good.  The results showed that the number of resonance peaks are very sensitive to the parameter $x$ and the 
thickness of the membrane.

For the gravitational field we show Fig. \ref{fig:gravity-x10-comdilaton}. In that picture
we can see the appearance of one peak of resonance for the parameters $x=1.0$ and $\lambda\sqrt{3M^3}=20$. However,
in Fig. \ref{fig:gravity-x15-comdilaton} we have the Transmission coefficient for gravity with dilaton for $x=1.5$ and $\lambda\sqrt{3M^3}=20$ presenting seven peaks of resonances. This shows the rule of the $x$ parameter 
in the setup chosen. We show for the scalar field, in Fig. \ref{fig:scalar-x10-semdilaton}, the transmission coefficient for analytical (lined) and numerical (dashed)  without dilaton with $x=1.0$, showing one peak of 
resonance. In Fig. \ref{fig:scalar-x10-comdilaton} with dilaton field we still have one peak of resonance for the parameters $x=1.0$ and $\lambda\sqrt{3M^3}=20$. In Fig. \ref{fig:scalar-x15-semdilaton} we have the 
Transmission coefficient for scalar without dilaton for $x=1.5$ and $\lambda\sqrt{3M^3}=20$ and, in Fig. \ref{fig:scalar-x15-comdilaton} showing eight peaks of resonances (the first one near $m=0$, can be made more visible 
if we just give a zoom in that region), the Transmission coefficient for scalar with dilaton for $x=1.5$ and $\lambda\sqrt{3M^3}=20$ showing three 
peaks. For the case of the vector field, we show in Fig. \ref{fig:vector-x10-comdilaton} and in Fig. \ref{fig:vector-x10-semdilaton}  the transmission coefficient for analytical (lined) and numerical (dashed) with and 
without dilaton with $x=1.0$ showing one peak of resonance, and in Fig. \ref{fig:vector-x15-comdilaton} we show the three peaks of the transmission coefficient for vector field with dilaton for $x=1.5$ and $\lambda\sqrt{3M^3}=20$

In the case of the Kalb-Ramond field we show in Fig. \ref{fig:tensor-x10-comdilaton} the transmission coefficient for analytical (lined) and numerical (dashed)  with dilaton with $x=1.0$ and 
in Fig. \ref{fig:tensor-x15-comdilaton} with $x=1.5$. In both cases we have $\lambda\sqrt{3M^3}=20$. In Fig. \ref{fig:tensor-x10-comdilaton} we see just one peak of resonance around $m=11.5$ and, 
in Fig. \ref{fig:tensor-x15-comdilaton} we have got five peaks. For the $q$-form case we give the graphics for the transmission coefficient in Fig. \ref{fig:tx10-comdil-alpha-1} and in Fig. \ref{fig:tx10-comdil-alpha-2} 
with $x=1.0$ to $\alpha=-1,-2$ with dilaton and in Fig. \ref{fig:tx10-semdil-c5} and Fig. \ref{fig:tx10-semdil-c7} with $p/2-q=5,7$ without dilaton. In the first one we can see two peaks of resonances, while in the 
remaining plots we have found just one peak. For completeness, we also give the graphics for the transmission coefficient  in Fig. \ref{fig:tx15-semdil-c5} and Fig. \ref{fig:tx15-semdil-c7} with $p/2-
q=5,7$ without dilaton. They show, respectively, six and eight peaks of resonances.

The transfer matrix method has been extensively used by the present authors to analyze resonances. But it was important to compare the numerical studies with 
the analytical ones to show that the numerical method is 
reliable and have a perfect agreement with solvable cases. In here we give this comparison for a lot of cases which encloses the presentation of the method. 
There are still more cases to be studied like, for example, 
splitting membranes, a case which would just change the Schr\"odinger potential. The study of fermionic resonances within this formalism is also a perspective of 
the present work. Another important aspect is related to corrections obtained, for example, to the Newton's law. Although this is not the main goal of the present work, it 
is important to mention that these corrections can be obtained through the transmission coefficient. The issue of metastable states has been better studied 
in Refs. \cite{Csaki:2000fc,Dvali:2000rv}. Generally an expression is given in the literature which involves $\psi_{m}(0)$, however, as 
stressed in \cite{Dvali:2000rv}, this is obtained after the spectral density expnasion in a complete set of modes. However, for our purpose it is better 
to keep the correction in terms of the spectral density, that is given by
\begin{equation}
 V(r)\sim \int \frac{e^{-my}}{r}\rho(m)dm.
\end{equation}

For a sharply peaked resonance we have $\rho=\delta(m-m_0)$ and we recover the well known formula for discrete masses \cite{Dvali:2000rv}. When there is a 
peak in $T$, by current conservation there is also a peak in the spectral density and the above integration can be approximated by a sum. From the knowledge 
of the spectral density other important quantities as the width of the resonance can also be obtained.

\section*{Acknowledgment}

We would like to acknowledge the 
financial support provided by Funda\c c\~ao Cearense de Apoio ao Desenvolvimento Cient\'\i fico e Tecnol\'ogico (FUNCAP), the Conselho Nacional de Desenvolvimento Cient\'\i fico e Tecnol\'ogico (CNPq) and FUNCAP/CNPq/PRONEX.

This paper is dedicated to the memory of my wife  Isabel Mara (R. R. Landim).

\end{document}